\documentclass[12pt]{article}
\pdfoutput=1
\usepackage{graphicx,psfrag,epsf,color}
\usepackage{xcolor}
\usepackage{amsmath,amssymb,amsfonts}
\usepackage{appendix}
\usepackage{multirow}
\usepackage{geometry}
\usepackage{float}
\usepackage{mathrsfs}  
\usepackage{authblk}
\usepackage{array}
\usepackage{cite}
\usepackage{slashed}
\usepackage{graphicx}
\usepackage{rotating}
\usepackage{slashed, cancel}
\usepackage[caption=false]{subfig}
\newcounter{MBQ}

\usepackage{tabularx}
\newcolumntype{C}{>{\centering\arraybackslash}X}

\newcommand{\eps}{\epsilon}

\def\be{\begin{equation}}
\def\ee{\end{equation}}
\def\beq{\begin{eqnarray}}
\def\eeq{\end{eqnarray}}
\newcommand{\bea}{\begin{eqnarray}}
\newcommand{\eea}{\end{eqnarray}}
\newcommand{\beas}{\begin{eqnarray*}}
\newcommand{\eeas}{\end{eqnarray*}}

\newcommand{\qboth}{\otimes}

\newcommand{\alem}{\alpha_{\rm em}}

\newcommand{\nn}{\nonumber}

\begin{document}
\allowdisplaybreaks

\begin{titlepage}

\begin{flushright}
{\small
TUM-HEP-1359/21\\
Nikhef-2021-018\\
August 09, 2021 
}
\end{flushright}

\vskip1cm
\begin{center}
{\Large \bf\boldmath Light-cone distribution amplitudes 
of light mesons \\[1ex] with QED effects}
\end{center}

\vspace{0.5cm}
\begin{center}
{\sc Martin~Beneke,$^a$ Philipp B\"oer,$^a$ 
Jan-Niklas Toelstede,$^{a}$ K. Keri Vos$^{b,c}$} \\[6mm]
{\it $^a$Physik Department T31,\\
James-Franck-Stra\ss{}e~1, 
Technische Universit\"at M\"unchen,\\
D--85748 Garching, Germany\\[0.3cm]

{\it $^b$Gravitational 
Waves and Fundamental Physics (GWFP),\\ 
Maastricht University, Duboisdomein 30,\\ 
NL-6229 GT Maastricht, the
Netherlands}\\[0.3cm]

{\it $^c$Nikhef, Science Park 105,\\ 
NL-1098 XG Amsterdam, the Netherlands}}
\end{center}

\vspace{0.6cm}
\begin{abstract}
\vskip0.2cm\noindent
We discuss the generalization of the leading-twist light-cone distribution amplitude for light mesons including QED effects. This generalization was introduced to describe virtual collinear photon exchanges above the strong-interaction scale $\Lambda_{\rm QCD}$ in the factorization of QED effects in non-leptonic $B$-meson decays. In this paper we study the renormalization group evolution of this non-perturbative function. For charged mesons, in particular, this exhibits qualitative differences with respect to the well-known scale evolution in QCD only, especially regarding the endpoint-behaviour. 
We analytically solve the evolution equation to first order in the electromagnetic coupling $\alem$, which resums large logarithms in QCD on top of a fixed-order expansion in $\alem$. We further provide numerical estimates for QED corrections to Gegenbauer coefficients as well as inverse moments relevant to (QED-generalized) factorization theorems for hard exclusive processes.
\end{abstract}
\end{titlepage}



\section{Introduction}
\label{sec:intro}
The study of QED effects in $B$-meson decays has become an active field of research in recent years~\cite{Beneke:2017vpq,Beneke:2019slt,Beneke:2020vnb,Beneke:2021jhp,Bordone:2016gaq,Isidori:2020acz,Mishra:2020orb}, as such effects will become increasingly more important when higher experimental precision is reached. The standard treatment of QED effects assumes the mesons to be point-like to very small distances of order $1/m_B$. However, this assumption neglects structure-dependent contributions; photons with wavelength of the order of, or smaller than the typical size $1/\Lambda_{\rm QCD}$ of a meson can resolve its partonic substructure. A way to systematically incorporate these effects is to disentangle the relevant energy scales below $m_B$ using effective field theories. This allows for a clear separation of very low-energetic photons with a point-like coupling to mesons from photons with energy of order $\Lambda_{\rm QCD}$ up to $m_B$, at which scale one can use the $1/m_B$ expansion. First studies following this path were made in~\cite{Beneke:2017vpq,Beneke:2019slt,Beneke:2020vnb,Beneke:2021jhp}. In particular, in~\cite{Beneke:2020vnb} we have shown that the QCD factorization formula \cite{Beneke:1999br,Beneke:2000ry} for non-leptonic charmless $B$ decays into two light mesons, $\bar{B} \to M_1 M_2$, can be extended to include QED corrections to all orders in $\alem$. Similar to the well-known QCD case, the relevant hadronic matrix elements can then be expressed as \cite{Beneke:2020vnb}\footnote{Contrary to~\cite{Beneke:2020vnb}, here we normalize to the QCD decay constant $f_M$ in the absence of QED, instead of introducing a QED-generalized decay constant $\mathscr{F}_M$.}
\begin{eqnarray}
\label{eq:QEDF}
\left\langle M_1 M_2 | Q_i |\bar B\right\rangle &=& \mathcal{F}^{BM_1}_{Q_2} \times  
T^{{\rm I}}_{i,Q_{2}} * f_{M_2} \Phi_{M_2} + \, T^{{\rm II}}_{i, \qboth} * f_{M_1} \Phi_{M_1} * f_{M_2}\Phi_{M_2} * f_{B} \Phi_{B,\qboth} \, .
\end{eqnarray}
This generalized factorization theorem expresses the matrix elements of weak effective operators $Q_i$ in the heavy-quark limit as convolutions of hard-scattering kernels $T_i$ with light-cone distribution amplitudes (LCDAs) $\Phi$ of heavy and light mesons. Besides the perturbatively calculable hard-scattering kernels discussed in detail in~\cite{Beneke:2020vnb}, a better understanding of adequately generalized non-perturbative objects is required. Therefore, the purpose of the present paper is to study the QED-generalized LCDA $\Phi_M(u;\mu)$ for light mesons $M = \pi, K, \ldots$. 

Compared to their definition in QCD-only, LCDAs for electrically charged mesons $M^\pm$ in QED exhibit some qualitatively new features which can be partly attributed to the non-decoupling of soft photons from the net charge of the meson, but also to the different electric charges of its constituents. First, this concerns the definition of a renormalizable function $\Phi_M(u;\mu)$, as its naive ultraviolet (UV) scale evolution is plagued by infrared (IR) sensitivity that needs to be removed by a ``soft rearrangement''. The thus properly defined LCDA can be used to derive its anomalous dimension and study the properties under scale variation using the renormalization group (RG). We find that the evolution kernel including QED effects contains, in addition to the QCD ERBL kernel~\cite{Lepage:1979zb, Lepage:1980fj,Efremov:1979qk}, new local logarithmic terms which have important consequences. For example, the evolution of Gegenbauer coefficients is no longer diagonal, and even the norm is no longer conserved, which spoils the interpretation of $\Phi_M(u;\mu)$ as a probability distribution. Although we do not aim at an analytic solution of the full QCD$\times$QED renormalization group equation (RGE), we provide numerical solutions and an analytic expression at $\mathcal{O}(\alem)$ that resums large logarithms in QCD on top of a fixed-order expansion in the electromagnetic coupling. 

QED corrections induce isospin-symmetry violation due to the different electric charges of up- and down-type quarks. For non-leptonic $B$ decays, this is of particular interest as it mimics short-distance electroweak penguin contributions which serve as a probe for new physics \cite{Beneke:2020vnb}. In the present work, we find that the distribution $\Phi_M(u;\mu)$ favours larger momenta of the $u$ quark, due to its larger electromagnetic coupling, resulting in a slightly asymmetric function even in the limit of massless quarks. This also modifies the endpoint behaviour of the LCDA, which eventually leads to an ill-defined evolution in the formal limit $\mu \to \infty$. Despite these new features, and contrary to the QED generalization of the $B$-meson LCDA~\cite{BLCDApaper}, the light-meson LCDAs (almost) retain their universality and are thus relevant to a variety of different hard exclusive processes. In this paper,  we study these new properties qualitatively, and estimate them quantitatively for realistic applications.

The outline of the paper is as follows.
In Section~\ref{sec:QEDERBL} we give the relevant basic definitions, including a brief review of the soft rearrangement required to define a renormalizable LCDA. Section~\ref{sec:ren} then states the evolution kernel, for which we study the endpoint behaviour of the LCDA in Section~\ref{subsec:pionEP}. The analytic first-order $\mathcal{O}(\alem)$ solution in Gegenbauer moment space is presented in Section~\ref{sec:geg}. We provide some numerical estimates of QED corrections to the Gegenbauer coefficients as well as inverse LCDA moments relevant in factorization theorems for hard exclusive processes in Section~\ref{subsec:pion:numestimates}. We conclude in Section~\ref{sec:conclusion}. We give technical details on the soft rearrangement and the endpoint behavior in two appendices.


\section{Basic definitions}
\label{sec:QEDERBL}
In QCD, light-cone distribution amplitudes for light mesons $M$ are well-established non-perturbative but universal objects, which appear in the theoretical description of hard exclusive processes at large energies.
They are defined as hadronic matrix elements of non-local operators composed of two light-like separated quark fields.
The twist-2 LCDA $\phi_M(u;\mu)$,
\begin{equation}
\label{eq:QCDtw2LCDA}
 \langle M(p)| \bar{q}_1(t n_+) [tn_+,0] \frac{\slashed{n}_+}{2}(1-\gamma_5) q_2(0)| 0\rangle  = \frac{i n_+p}{2}\int_0^1 du\, e^{i u (n_+p) t}  f_M \phi_{M}(u;\mu) \ ,
\end{equation}
is usually the leading contribution in the twist expansion.
In the above definition, the light-like reference vectors $n_\pm^\mu$, obeying $n_\pm^2=0, n_+\cdot n_-=2$, are conveniently defined through the meson's momentum \mbox{$p^\mu = E n_-^\mu + m_M^2/(4E) \, n_+^\mu$} in the frame where $E \gg m_M$ is the large energy of order the hard scale of the process. The function $\phi_M(u;\mu)$ is, however, boost-invariant. The displaced fields are connected by a straight light-like Wilson line $[tn_+,0]$ of finite length to ensure gauge invariance of the definition in \eqref{eq:QCDtw2LCDA}. 
It is sometimes convenient to express the finite-distance Wilson line in terms of infinite Wilson lines as $[tn_+,0] = W(tn_+)W^\dagger(0)$, with
\begin{equation}
    W(x) = {\mathbf P} \exp \left\{ i g_s \int_{-\infty}^0 d s \, n_+ G(x + s n_+) \right\} \, .
 \label{eq:WdefQCD}
\end{equation}
Finally, $f_M$ is the scale-independent meson decay constant in QCD, defined through the local limit $t=0$ of~\eqref{eq:QCDtw2LCDA}.
This implies the normalization condition $\int_0^1 du\, \phi_M(u;\mu) = 1$.

In the present work, we study the QED generalization $\Phi_M(u;\mu)$ of the leading-twist LCDA $\phi_M(u;\mu)$. By this we mean that the matrix element is computed with $\mathcal{L}_{\rm QCD+QED}$, which accounts for an arbitrary number of virtual collinear photon exchanges between the (electrically charged) constituents of $M$ on top of the strong interaction.
For neutral mesons $M^0$, one has $Q_{q_1}=Q_{q_2}=Q_q$, and the definition~\eqref{eq:QCDtw2LCDA} remains valid after modifying the Wilson line to include the photon field $A^\mu$:
\begin{equation}
    W^{(q)}(x) = \exp \left\{i Q_q e \int_{-\infty}^0 d s \, n_+ A(x + s n_+) \right\}  \, 
 {\mathbf P} \exp \left\{ i g_s \int_{-\infty}^0 d s' \, n_+ G(x + s' n_+) \right\} \, ,
 \label{eq:Wdef}
\end{equation}
where $Q_q$ denotes the electric charge of the quark field $q$ in units of $e=\sqrt{4\pi \alem}$.
The situation is, however, different for electrically charged mesons $M^\pm$.
For definiteness, we consider $q_1 = D = d,s$ and $q_2=u$, such that $M$ has total charge \mbox{$Q_M = Q_{q_1}-Q_{q_2}= -1$}.
The corresponding case for $M^+$ is related by CP invariance of QCD and QED.
The gauge-invariant bilinear non-local operator now takes the form
\begin{align}
\label{eq:QEDchargedMop}
    \bar{D}(t n_+) W^{(d)}(tn_+) \frac{\slashed{n}_+}{2}(1-\gamma_5) W^{\dagger(u)}(0) u(0) \,.
\end{align}
We first notice that gauge invariance dictates the operator to extend on the infinite light-ray, instead of being localized on a finite interval $[t n_+,0]$.
This can be seen by combining the Wilson lines associated with the quark fields of different electric charge to $W^{(d)}(tn_+) W^{\dagger(u)}(0) = [tn_+,0]^{(d)} W^{(Q_M)}(0)$. In addition to the gauge-link $[tn_+,0]^{(d)}$ associated with the charge $Q_d$, the operator contains a Wilson line extending from $-\infty$ to $0$ with the total electric charge of the meson $Q_M = Q_d - Q_u$, defined as 
\begin{equation}
    W^{(Q_M)}(x) \equiv \left(W^{(d)} W^{\dagger(u)}\right)(x) = \exp \left\{i Q_M e \int_{-\infty}^0 d s \, n_+ A(x + s n_+) \right\}  ,
\end{equation}
where the QCD part of the Wilson line cancels in $W^{(Q_M)}(x)$ due to the meson $M^-$ being a colour-singlet. 

As discussed in detail in~\cite{Beneke:2019slt,Beneke:2020vnb}, due to the non-decoupling of soft photons from electrically charged mesons, the operator~\eqref{eq:QEDchargedMop} itself is no longer renormalizable in the sense that its anomalous dimension is IR divergent. This is due to a non-trivial overlap between the soft and collinear sector in QED-generalized collinear factorization theorems. This overlap renders the UV divergences of the collinear part of the operator dependent on the IR regulator. To remove this overlap and make the operator renormalizable, it is sufficient to multiply the collinear operator with a remnant of the soft function of the process which must be present if the entire process is to be IR safe. 
One particular choice---inspired by the soft function for a decay of a neutral particle into two back-to-back charged particles---is to define subtraction factors $R_c$ and $R_{\bar{c}}$ through the following vacuum matrix element of soft Wilson lines:
\begin{equation}
\left| \langle 0 |\big(S_{n_-}^{\dagger (Q_{M})} S_{n_+}^{(Q_{M})}\big)(0)\,|0\rangle\right| \equiv R_{c}^{(Q_{M})}  R_{\bar{c}}^{(Q_{M})} \;.
\label{eq:softsubtraction}
\end{equation}
The soft Wilson lines originate from the coupling of soft photons to the electrically charged constituents of the particles in the process. For outgoing antiquarks with electric charge $Q_q$, the soft Wilson line reads
\begin{equation}
\label{eq:softwilsonlines}
 S^{(q)}_{n_\pm}(x) = \exp \left\{ -i Q_q e \int_0^{\infty} d s \, n_\pm A_{s}(x + s n_\pm) \right\}  \, 
 {\mathbf P} \exp \left\{ -i g_s \int_0^{\infty} d s' \, n_\pm G_s(x + s' n_\pm) \right\} \,,
\end{equation}
while $S^{\dagger(q)}_{n_\pm}$ must be used for outgoing quarks with electric charge $Q_q$. As for the collinear case, the soft Wilson line depending on the total electric charge $Q_M$ in \eqref{eq:softsubtraction} is defined by
\begin{equation}
    S^{(Q_M)}_{n_\pm}(x) \equiv \left(S_{n_\pm}^{(d)} S_{n_\pm}^{\dagger(u)}\right)(x) = \exp \left\{ -i Q_M e \int_0^{\infty} d s \, n_\pm A_{s}(x + s n_\pm) \right\} \; .
\end{equation}
The factors $R_{c}^{(Q_{M})}$ and $R_{\bar{c}}^{(Q_{M})}$ are defined such that their UV divergences only depend on the IR-regulator associated with the corresponding collinear direction. This cancels the regulator-dependent terms in the anomalous dimension of the collinear operator, leaving a renormalizable operator.
We have taken the absolute value of the matrix element~\eqref{eq:softsubtraction} to avoid spurious imaginary terms due to soft rescattering phases in the collinear sector.
More details and explicit expressions for the subtraction factors in an off-shell regularization scheme can be found in Appendix~\ref{app:sec:notation} as well as the next subsection.
Hence, we define the QED-generalized LCDA for an electrically charged light meson as
\begin{equation}
\label{eq:M2LCDA}
 \langle M^-(p)| R_c^{(Q_M)} \, 
(\bar{D} W^{(d)}) (tn_+) \frac{\slashed{n}_+}{2}(1-\gamma_5) (W^{\dagger(u)} u)(0)| 0\rangle  = \frac{i n_+p}{2}\int_0^1 du\, e^{i u (n_+p) t}  f_M \Phi_{M}(u;\mu) \, .
\end{equation}
We emphasize that we \emph{choose} to normalize $\Phi_M(u;\mu)$ with respect to the renormalization-scale independent QCD decay constant $f_M$ in the absence of QED.
We do not pull out the local limit of the operator in QED, since it would mix into higher logarithmic moments under renormalization group evolution. Hence, also the normalization condition is no longer fulfilled at any scale $\mu$, $\int_0^1 du\, \Phi(u;\mu) \neq 1$.

The QED generalized LCDA defined in~\eqref{eq:M2LCDA} is
renormalizable and has a well-defined UV scale evolution. It appears as part of the factorization formula~\eqref{eq:QEDF} for the IR divergent non-radiative amplitude that describes virtual photon (and gluon) interactions above the strong interaction scale $\Lambda_{\rm QCD}$. The complete process is IR finite once real photon radiation of undetected photons with energies below a sufficiently small resolution $\Delta E$ is accounted for. The above LCDA is relevant when the energy of the radiated photons is much smaller than $\Lambda_{\rm QCD}$ (``ultrasoft''), or of the order of $\Lambda_{\rm QCD}$ (``soft''), in a frame where the electrically charged particles are ultrarelativistic. The LCDAs are defined through exclusive matrix elements and are hence themselves IR divergent. Their proper non-perturbative definition contains a prescription for subtracting these divergences. In the complete description of the process they then appear as IR-finite matching coefficients for the effective theory
of ultrasoft radiation. In the application to non-leptonic decays
\cite{Beneke:2020vnb}, the matching was performed in dimensional regularization,
and hence the IR divergences in the LCDA were minimal
subtracted, resulting in an IR-subtraction scale dependence,
which is {\em not} the subject of the present paper (see
also discussion at the end of Sec. 5.2. of \cite{Beneke:2020vnb}).

\begin{figure}[!t]
    \centering
    \includegraphics[scale=1.2]{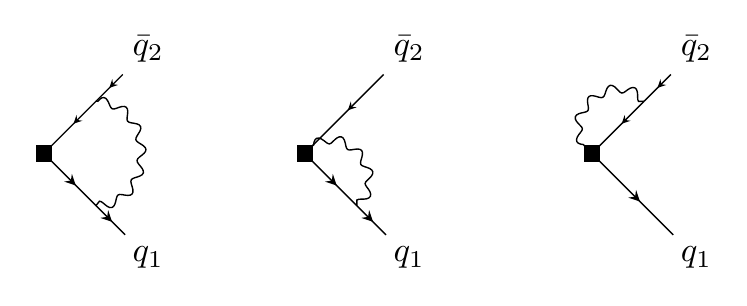}
    \caption{Diagrams that contribute to the renormalization of the operator~\eqref{eq:QEDchargedMop} at $\mathcal{O}(\alem)$.}
    \label{fig:QEDERBL}
\end{figure}

An important remark is to be made about the universality of $\Phi_M(u;\mu)$. Although we defined the subtraction factors $R_c$ and $R_{\bar{c}}$ in \eqref{eq:softsubtraction} by implicitly assuming one additional anti-collinear direction $n_+^\mu$, $\Phi_M(u;\mu)$ remains a universal object relevant to various two- and multi-body exclusive processes. This can be seen by performing the soft rearrangement for a generic $n$-jet SCET operator: \begin{align}
\label{eq:njetrearrangement}
    \mathcal{O}_{\rm eff} = \mathcal{O}_{s} \times \mathcal{O}_{c_1} \mathcal{O}_{c_2} \dots \mathcal{O}_{c_n} = \frac{\mathcal{O}_{s}}{R_{c_1} R_{c_2} \dots R_{c_n}} \times (R_{c_1} \mathcal{O}_{c_1}) (R_{c_2} \mathcal{O}_{c_2}) \dots (R_{c_n} \mathcal{O}_{c_n}) \,.
\end{align}
For every light-like direction $n_{i-}$ one defines a corresponding back-to-back vector $n_{i+}$ with $n_{i+} n_{i-} = 2$. The vector $n_{i+}$ does not have to coincide with the direction of flight of any particle involved in the process. Then the subtraction factors $R_{c_i}$ and $R_{\bar{c}_i}$ are defined through vacuum matrix elements of two soft Wilson lines along $n_{i-}$ and $n_{i+}$ as in~\eqref{eq:softsubtraction}. The so-defined collinear sectors are individually renormalizable and so is the left-hand side of~\eqref{eq:njetrearrangement}. Thus, by consistency of SCET as an effective theory, also the rearranged soft operator is renormalizable. The definition implies the choice of a ``soft reference frame'', which for $B$ meson decays is naturally the $B$ meson rest frame. An unavoidable consequence of the QED effects is the breaking of boost invariance in the LCDA definition, which results in a dependence of $\Phi_M(u;\mu)$ on the large energy $E$ of the meson measured in the soft reference frame.

Finally, we remark that once isospin breaking effects are considered, neutral $\pi$ mesons are described by two distinct quark LCDAs. These can either be defined by $\Phi_{M^0}^{(u)}$ and $\Phi_{M^0}^{(d)}$, or alternatively as the $SU(2)$ singlet and triplet linear combinations. In general, the different LCDAs for neutral mesons will mix under renormalization, including a two-gluon LCDA. 
We do not compute this mixing in this paper, because we are mainly interested in electrically charged mesons, where the non-decoupling of soft photons is important. 

\section{Renormalization}
\label{sec:ren}
The renormalized non-local operator $\mathcal{O}^{\rm ren}(u;\mu)$ is related to the bare operator through
\begin{equation}
    \label{eq:pionZfactor}
    \mathcal{O}^{\rm ren}(u;\mu) = \int_0^1 dv \, Z(u,v;\mu) \mathcal{O}^{\rm{bare}} (v)  \; .
\end{equation}
To derive the $Z$-factor of the operator in~\eqref{eq:QEDchargedMop}, we calculate the diagrams in Fig.~\ref{fig:QEDERBL} with slightly off-shell external quark states, and add the $\overline{\rm{MS}}$ renormalization factors for the quark fields. 

The resulting UV poles in dimensional regularization with space-time dimensions $d=4-2\epsilon$ have already been given in (21) and (22) in \cite{Beneke:2020vnb}.
For completeness, we repeat the $\mathcal{O}(\alem,\alpha_s^0)$ result (with slightly adapted notation)
\begin{eqnarray}
\label{eq:UVpolescbar}
Z(u,v) &=& \delta(u-v)- \frac{\alem}{4\pi}\bigg\{ Q_M \delta(u-v) \left( Q_M \left(\frac{2}{\epsilon^2} + \frac{3}{2\eps} \right) + \frac{2Q_{q_1}}{\eps}  \ln\frac{\mu^2}{-k_{q_1}^2} - \frac{2Q_{q_2}}{\eps}  \ln \frac{\mu^2}{-k_{q_2}^2} \right) \nonumber  \\ 
&&\hspace*{-1.5cm}+ \,\frac{2 Q_{q_1} Q_{q_2}}{\eps} \left[ \left(1+\frac{1}{v-u}\right) \frac{u}{v}\, \theta(v-u) + \left(1+\frac{1}{u-v}\right) \frac{\bar{u}}{\bar{v}} \,\theta(u-v) \right]_+^{(u)} \bigg\} + \mathcal{O}(\alpha_s,\alem^2) \,, \quad
 \end{eqnarray} 
where $\bar{x}\equiv 1-x$ and for simplicity we dropped the $\mu$ in the argument of $Z$. Here and below, $\alem (\alpha_s)$ denotes the electromagnetic (strong) coupling in the $\overline{\rm MS}$ scheme at the scale $\mu$. The plus-distribution is defined in the variable $u$:
\begin{equation}
 \int_0^1 du \, \Big[ \dots \Big]_+^{(u)} f(u) \equiv
 \int_0^1 du \, \Big[ \dots \Big] (f(u) - f(v))\,.
\end{equation}
As discussed earlier, the UV divergences depend on the off-shellness of the external quark states, $k_{q_1}^2$ and $k_{q_2}^2$, inconsistent with renormalization. This dependence is removed by multiplying with $R_c$, which at $\mathcal{O}(\alem)$ reads:
\begin{equation}
\label{eq:Rcbar}
R_c^{(Q_{M})} = 1 -  \frac{\alem}{4\pi} Q_{M}^{\,2} \left[\frac{1}{\epsilon^2} + \frac{2}{\epsilon}\ln{\frac{\mu}{-\delta_c}} + \mathcal{O}(\epsilon^0) \right] .
\end{equation}
Here $\delta_c$ is the remnant of the off-shell regulator in the soft matrix element (see Appendix~\ref{app:sec:notation} for more details). Since soft Wilson lines obey the multiplication law $S^{\dagger (q_1)}_{n_-} S^{(q_2)}_{n_-} =S^{\dagger (Q_M)}_{n_-}$, the constraint $\delta_c = k_{q_1}^2/(n_+k_{q_1}) = k_{q_2}^2/(n_+k_{q_2})$ must be imposed on a consistent off-shell regularization prescription.

After this rearrangement the operator is renormalizable and we can compute its anomalous dimension via
\begin{align}
\label{eq:ADdef}
    \Gamma(u,v;\mu) = - \int_0^1 dw \frac{d Z(u,w;\mu)}{d \ln \mu} Z^{-1}(w,v;\mu) \; .
\end{align}
Then $\Phi(u;\mu)$ obeys the renormalization group equation (RGE)
\begin{equation}
\label{eq:pionRG}
    \frac{d}{d \ln \mu} \Phi_M(u;\mu) = - \int_0^1 dv~\Gamma(u,v;\mu) \Phi_M(v;\mu) \; .
\end{equation}
From the one-loop $Z$-factor in~\eqref{eq:UVpolescbar}, including now the standard one-loop QCD renormalization, after multiplication with~\eqref{eq:Rcbar}, we obtain from~\eqref{eq:ADdef} the one-loop anomalous dimension
\begin{align}
\label{eq:ERBL}
  \Gamma(u,v;\mu) = &- \frac{\alpha_s C_F + \alem Q_{q_1} Q_{q_2}}{\pi} \left[ \left(1+\frac{1}{v-u}\right) \frac{u}{v} \theta(v-u) + \left(1+\frac{1}{u-v}\right) \frac{\bar{u}}{\bar{v}} \theta(u-v) \right]_+^{(u)} \nn  \\ &- \frac{\alem}{\pi} \delta(u-v) Q_M\left( Q_M \left( \ln \frac{\mu}{2E} + \frac34 \right) - Q_{q_1} \ln {u} + Q_{q_2} \ln \bar{u} \right) .
\end{align}
Note that this result holds for both charged ($q_1=d, q_2 = u, Q_M = -1$) and neutral ($Q_{q_1}=Q_{q_2}=Q_q, Q_M = 0$) mesons. In the latter case the local contributions in the second line of~\eqref{eq:ERBL} vanish and we recover the standard ERBL evolution kernel~\cite{Lepage:1979zb, Lepage:1980fj,Efremov:1979qk} with a modified one-loop coefficient $\alpha_s C_F \to \alpha_s C_F+\alem Q_q^2$.
For charged mesons, however, the second line in~\eqref{eq:ERBL} implies additional logarithmically enhanced local terms. In particular, the anomalous dimension now depends on the large energy $E$ of the meson, which is the hard scale of the process. It is important to emphasize that this energy is measured in a frame that defines the soft modes of the process, i.e.~a frame in which the meson is ultrarelativistic. This can be viewed as a manifestation of the factorization anomaly \cite{Beneke:d2005} (or collinear anomaly~\cite{Becher:2010tm}) in SCET, which requires the explicit breaking of boost invariance, in this case through the soft rearrangement.
The energy dependence is independent of the meson constituents and only related to its overall charge, which can be seen by comparing to the corresponding collinear matrix element $Z_\ell(\mu)$ for a point-like fermion with electric charge $Q_\ell = Q_M$ and energy $E_\ell = E$. The precise definition of $Z_\ell$ as well as its one-loop UV divergences can be found in equations~(62) and (63) of~\cite{Beneke:2020vnb}.
Its anomalous dimension and RG evolution reads \begin{equation} \label{eq:Zell}
    \frac{d}{d\ln \mu} Z_\ell(\mu) =  \frac{\alem}{\pi} Q_{\ell}^2 \left( \ln \frac{\mu}{2E} + \frac{3}{4} \right) Z_\ell(\mu) \; ,
\end{equation}
with solution $Z_\ell(\mu)= U_\ell(\mu,\mu_0) Z_\ell(\mu_0)$. The multiplicative evolution factor\footnote{We emphasize that, according to the definition in~\cite{Beneke:2020vnb}, the universal double-logarithmic part of this evolution factor is not part of non-radiative amplitude.} is
\begin{align}
\label{eq:uell}
    U_\ell(\mu,\mu_0) = \exp \bigg\{ \int_{\mu_0}^\mu \frac{d\mu'}{\mu'} \frac{\alem(\mu')}{\pi} Q_{\ell}^2 \bigg( \ln\frac{\mu'}{2E} + \frac{3}{4} \bigg) \bigg\} \; .
\end{align}
We note that in the QED-generalized factorization theorem for non-leptonic $B$-meson decays the ratio $\Phi_M(u)/Z_\ell$ naturally appears \cite{Beneke:2020vnb}, and we will thus normalize the LCDA accordingly in Sec.~\ref{sec:geg}. The explicit quark charges in the second line of~\eqref{eq:ERBL} violate isospin symmetry and hence the new local terms render the LCDA asymmetric in $u \leftrightarrow 1-u$. Further, an expansion in Gegenbauer polynomials will no longer diagonalize the kernel. 

\section{Endpoint behaviour}
\label{subsec:pionEP}
In this section, we study how the QED contributions to the anomalous dimension change the well-known linear endpoint behaviour of the light-meson LCDAs in QCD. The linear endpoint behavior can be inferred from conformal symmetry arguments, see e.g.~\cite{Braun:2003rp}. This approach uses a construction of conformal operators at the RG fixed point, where the QCD $\beta$-function vanishes, which dictates the form of the anomalous dimension. In QCD$\times$QED there is no related RG fixed point since the $\beta$-functions depend on both $\alpha_s$ and $\alem$. Therefore, the anomalous dimension is not restricted by these arguments. The following analysis 
of the endpoint behaviour is based on the one-loop kernel and 
may not apply at (uninterestingly) large scales, when the 
QED coupling becomes strong. 

To study solutions of the RGE, it is useful to rewrite the plus-distribution in the anomalous dimension~\eqref{eq:ERBL} as a distribution in the variable $v$: 
\begin{align}
\label{eq:pdinv}
 \Big[ \dots \Big]_+^{(u)} = \Big[ \dots \Big]_+^{(v)} + \left(u \ln \bar{u} + \bar{u} \ln u + \frac32 \right) \delta(u-v) \,. 
\end{align}
We then split the integral in $v$ in the RGE into different momentum regions in order to construct an asymptotic expansion of $\Phi_M(u;\mu)$ near the endpoints $u=0$ and $u=1$. Throughout the rest of this section we focus on the limit $u\to 0$; the behaviour near $u\to 1$ is qualitatively similar and follows from the replacements of charge factors, $Q_{q_1} \leftrightarrow Q_{q_2}$ (which implies $Q_M \to -Q_M$). For small $u$, the integral over $v$ in \eqref{eq:pionRG} receives contributions from two regions: the soft region, $v \sim u \ll 1$, and a ``true'' collinear region, $v \sim 1, u \ll 1$. In these regions the integral of the plus-distribution part simplifies to
\begin{align}
    &\text{Collinear region:} &\int_0^1 dv\Big[ \dots \Big]_+^{(v)} &\to u \int_0^1 \frac{dv}{v} \left(1+\frac1v \right) \; , \label{eq:ERBLregioncol} \\
    &\text{Soft region:} &\int_0^1 dv\Big[ \dots \Big]_+^{(v)} &\to u \int_0^\infty dv \, \Big[ \frac{\theta(v-u)}{v(v-u)} + \frac{\theta(u-v)}{u(u-v)} \Big]_+^{(v)} \; \label{eq:ERBLregionsoft} .
\end{align}
Based on power counting, the plus distribution has been omitted in the collinear region, while the expansion in the soft region implies that the evolution kernel acts on a function space with support on the whole positive real axis $[0,\infty)$. After these simplifications, the integrals in the collinear and soft region may become divergent for $v\to 0$ and $v\to \infty$, respectively, and need to be regularized. However, in the expansion by regions~\cite{Beneke:1997zp}, the integrand determines the power-counting, and we can neglect regions that give a suppressed contribution irrespective of whether the integral converges or not.

In order to determine which of the two regions dominates we assume that $\Phi_M(u;\mu_0) \sim u^b$ for the small-$u$ behaviour at the initial scale $\mu_0$. 
We then distinguish the following cases:
\begin{itemize}
    \item For $b>1$ the integral \eqref{eq:pionRG} is dominated by the collinear region which scales as $u^1$. The soft region as well as the local terms in the RGE count as $u^b \ll u^1$ and can be dropped. Hence, an infinitesimal evolution $\mu_0 \to \mu_0 + d\mu$ generates a term $\Phi_M(u;\mu_0+d\mu) \sim u^1$ which now dominates over $u^b$. It follows that, for $b>1$, RG evolution in the collinear region always drives the LCDA immediately back to linear asymptotic behaviour.
    \item For $b=1$ the soft and the collinear region contribute equally and there is no apparent simplification of the evolution kernel. We will argue below that we can nevertheless determine the asymptotic form of $\Phi_M(u;\mu)$ from the soft approximation.
    \item For $b<1$ the soft region dominates and is of the same power $u^b$ as the local terms in the RGE (expanded for $u \to 0$). This implies that the endpoint behaviour of $\Phi_M(u;\mu_0+d\mu)$ is now fully determined by the endpoint behaviour of $\Phi_M(u;\mu_0)$. 
    \item Lastly, for $b\leq -1$ the convolution integral in the RGE \eqref{eq:pionRG} is ill-defined.
    Interestingly, for charged mesons we find that RG evolution to extremely large scales
    inevitably drives the solution to this scenario due to the local terms and the coefficient of the ERBL kernel in~\eqref{eq:ERBL}.
\end{itemize}
To make the discussion more transparent it is instructive to first analyze the well-known evolution via the QCD-only one-loop kernel along these lines.
Assuming an initial condition at the scale $\mu_0$ with $b<1$, the asymptotic soft evolution kernel takes the form 
\begin{align}
\label{eq:LNtypekernel}
  \Gamma(u,v)\big\vert_{\text{soft}, \alem =0} = &- \frac{\alpha_s C_F}{\pi} \left\{\left[ \frac{\theta(v-u)}{v(v-u)} + \frac{\theta(u-v)}{u(u-v)}  \right]_+^{(v)} u + \delta(u-v) \left(\ln u+\frac32 \right) \right\} \,.
\end{align}
Interestingly, up to a constant in the local part, this precisely reproduces the evolution kernel~\cite{Lange:2003ff} for the $B$-meson LCDA $\phi_B^+(\omega)$.
This is intuitive, since for $u \to 0$ the $D$ quark in the $M^-$ has only soft fluctuations and the large momentum component $n_+ k_{\bar{u}}$ of the anti-$u$ quark  becomes frozen. To analyze the asymptotic behaviour we make use of some techniques developed in~\cite{Bosch:2003fc,Liu:2020eqe}.
It turns out to be convenient to study the RG evolution in Mellin space:
\begin{align}
\label{eq:mellintrafo}
 \tilde{\Phi}_M(\eta;\mu) \equiv \int_0^\infty du \, u^{-1-\eta} \,\Phi_M(u;\mu) \,, \qquad \Phi_M(u;\mu) = \int_{c-i\infty}^{c+i\infty} \frac{d\eta}{2\pi i} \, u^{\eta} \,\tilde{\Phi}_M(\eta;\mu) \,,
\end{align}
where $c$ is a real parameter. The definition obviously holds also for the QCD-only LCDA $\phi_M(u;\mu)$. If we assume some $\mu$-dependent exponent $\Phi_M(u;\mu)\sim u^{b_\mu}$ for $u\to 0$, the Mellin transform converges for ${\rm Re}(\eta)<b_\mu$, so that $c<b_\mu$ must be chosen. Without QED, the Mellin-space RGE is
\begin{align}
 \left[\frac{d}{d\ln\mu} + \frac{\alpha_s C_F}{\pi} \partial_\eta \right] \tilde{\phi}_M(\eta;\mu) = - \frac{\alpha_s C_F}{\pi} \left(H_\eta + H_{-\eta} -\frac32\right) \tilde{\phi}_M(\eta;\mu) \,,
\end{align}
where $H_n = \gamma_E + \psi(n+1)$, with $\psi(n) = \Gamma'(n)/\Gamma(n)$ the digamma function, is the harmonic number function. This equation is solved by \cite{Bell:2013tfa}
\begin{align}
\label{eq:mellinsolutionQCD}
\tilde{\phi}_M(\eta;\mu) &= e^{(2 \gamma_E -3/2) a}  \,\frac{\Gamma(1-\eta) \Gamma(1+\eta+a)}{\Gamma(1+\eta)\Gamma(1-\eta-a)} \,\tilde{\phi}_M(\eta+a;\mu_0) \, .
\end{align}
The Mellin variable $\eta$ of the initial condition on the right-hand side shifted by the evolution variable
\begin{align} \label{eq:aQCD}
 a \equiv a(\mu,\mu_0) = - C_F \int_{\mu_0}^\mu \frac{d\tilde{\mu}}{\tilde{\mu}} \frac{\alpha_s(\tilde{\mu})}{\pi} = \frac{2C_F}{\beta_0^{\rm QCD}} \ln \frac{\alpha_s(\mu)}{\alpha_s(\mu_0)} + \mathcal{O}(\alpha_s) \,,
\end{align}
which is always negative for evolution to higher scales, that is, $a<0$ for $\mu>\mu_0$. Here the one-loop QCD $\beta$ function is
\begin{equation}
\label{eq:betafcts}
    \frac{d\alpha_s}{d\ln \mu} = -  \frac{\alpha_s^2}{2\pi}\beta_0^{\rm QCD}  \qquad\mbox{with}\qquad\beta_0^{\rm QCD}=\frac{11}{3} N_c -\frac23 n_f \; , 
\end{equation}
where $N_c = 3$ and $n_f$ is the number of active quark flavours.

Eqs.~\eqref{eq:mellintrafo} and \eqref{eq:mellinsolutionQCD} require the contour parameter $c$ to lie in the interval $-1-a < c< b_\mu \equiv b-a(\mu,\mu_0)$ such that the integral converges.  
The asymptotic behaviour of $\phi_M(u;\mu)$ for small $u$ is then determined by the location of the left-most pole on the real axis for ${\rm Re}(\eta)>c$ of the solution~\eqref{eq:mellinsolutionQCD} after closing the integration contour in the right half-plane. The initial condition $\phi(u;\mu_0)\sim u^b$ implies that its Mellin transform has a singular point at $\eta = b$ and thus, the shifted function in~\eqref{eq:mellinsolutionQCD} at $\eta = b_\mu$. As long as $b_\mu <1$, the analytic structure of $\tilde{\phi}_M(\eta+a;\mu_0)$, rather than the $\Gamma(1-\eta)$ in the prefactor in \eqref{eq:mellinsolutionQCD}, determines the asymptotic behaviour of the evolved function to be $\phi_M(u;\mu) \sim u^{b_\mu}$, see Appendix B for more details. The precise functional form, i.e. whether or not this power-law is modified by additional powers of $\ln u$, depends on the initial condition. For the rest of this section we restrict ourselves to pure power-like initial conditions. With increasing $\mu$ one will eventually reach the point $b_\mu =1$. (Recall that for exponents greater than one the collinear region dominates and immediately generates a linear term.) Due to the gamma function $\Gamma(1-\eta)$ in the Mellin-space solution~\eqref{eq:mellinsolutionQCD}, one might conclude that $\phi_M(u;\mu) \sim u^1$.
This is true, although the asymptotic kernel~\eqref{eq:LNtypekernel} is not the right object to begin with, since the collinear region is of the same order in power-counting as the soft region. As this collinear region does not contribute additional $\ln u$ enhanced terms, the soft region produces the correct asymptotic form of the LCDA. This becomes apparent from the expressions in~\eqref{eq:ERBLregioncol}. The integral in the soft region in \eqref{eq:ERBLregionsoft}, on the other hand, does contribute such terms due to the logarithmic integral $\int_u^\infty \frac{dv}{v^2} \phi_M(v;\mu)$ for $\phi_M(v;\mu) \sim v^1$.

\begin{figure}[t]
\centering
\includegraphics[width=0.49\textwidth]{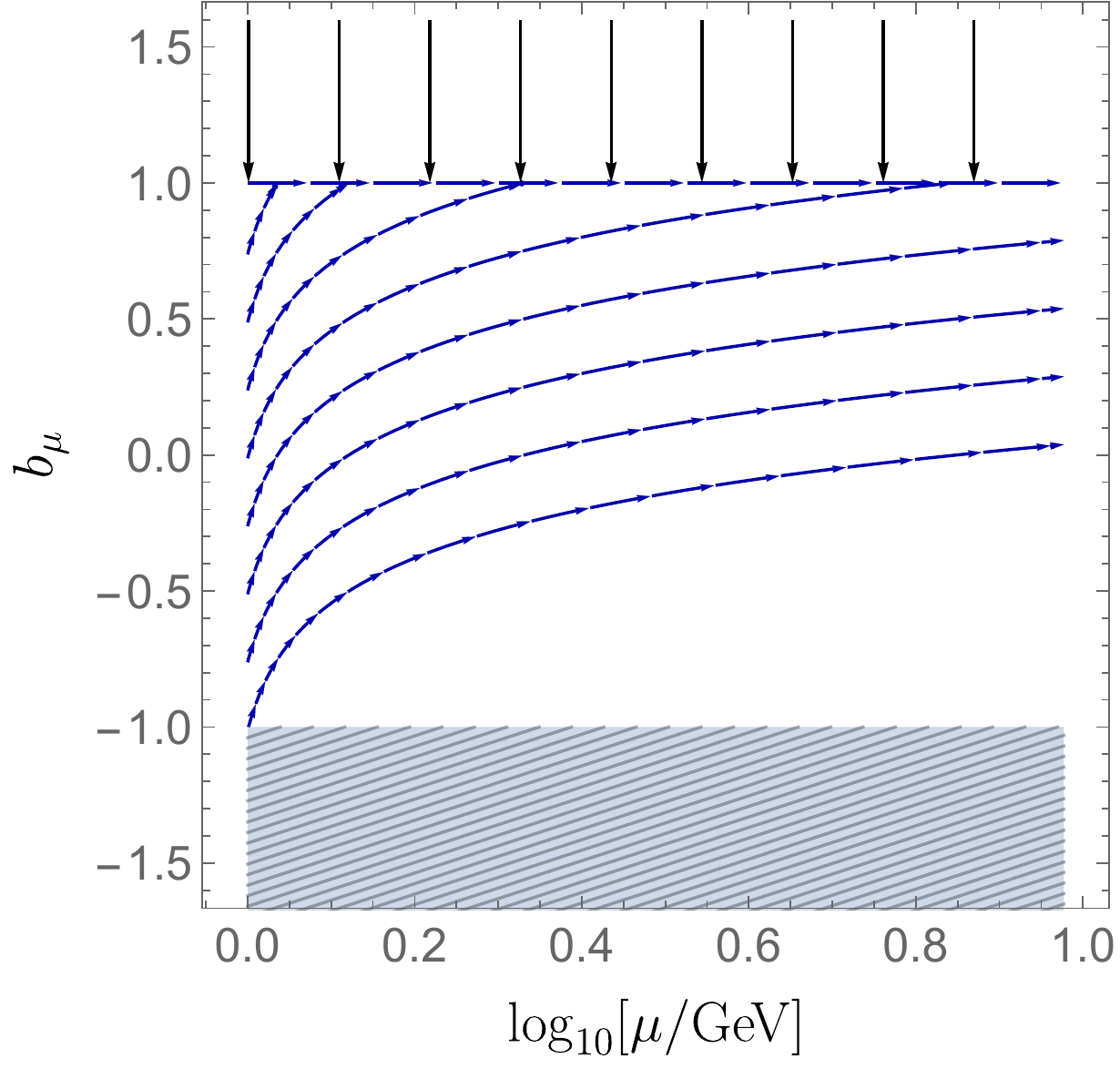}
\includegraphics[width=0.49\textwidth]{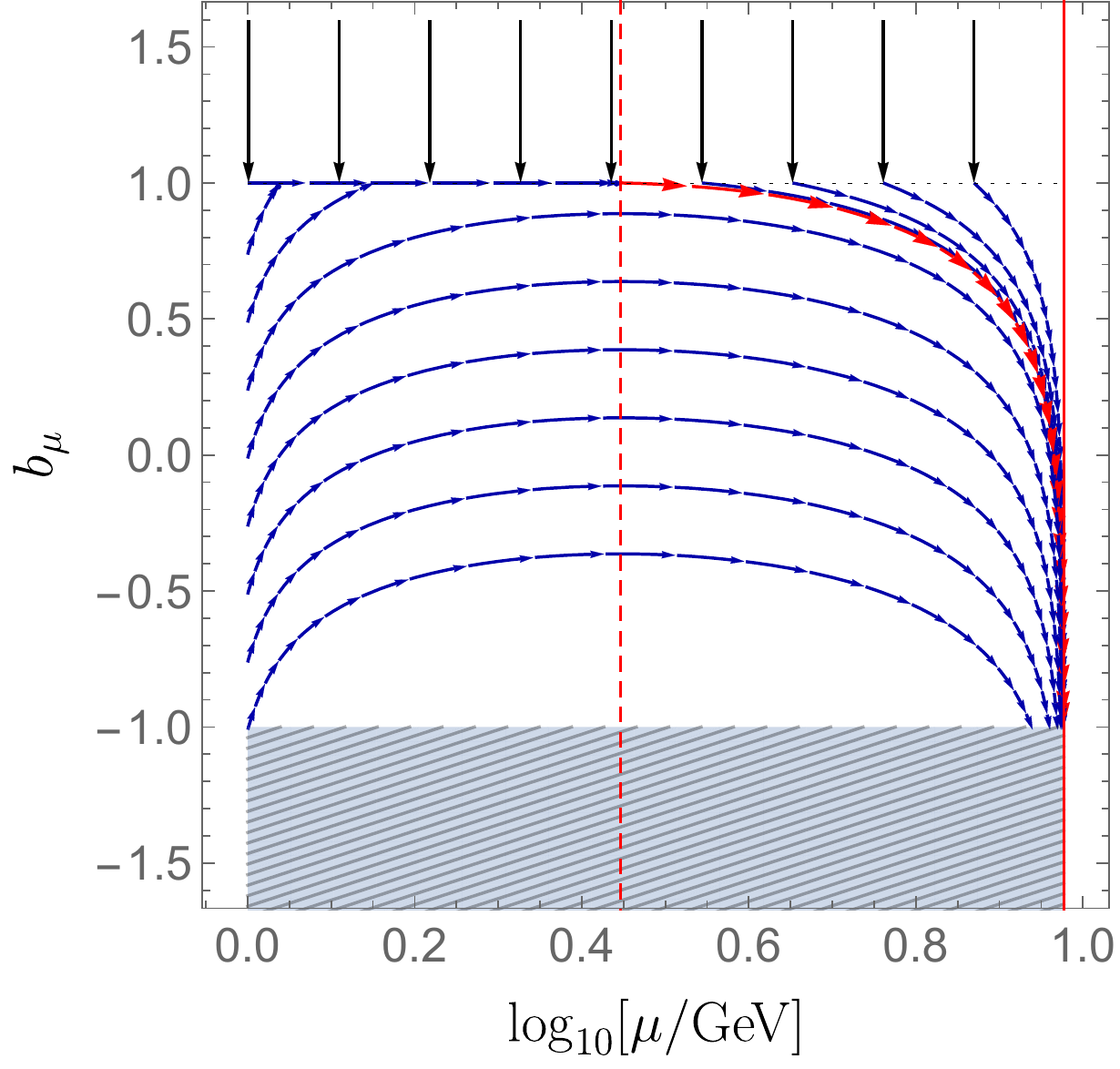} 	
\caption{RG flow of the exponent $b_\mu$ in QCD-only (left), and QCD$\times$QED (right). For illustration purposes, these curves are obtained with $\alpha_s(\mu_0)=4\pi$, $\alem(\mu_0)=\pi/4$ at the initial scale $\mu_0 = 1 \text{ GeV}$, with $\beta_0^{\rm QCD} = 29/3$ and $\beta_0^{\rm QED} = -32/9$, which corresponds to a theory with one generation of quarks and leptons. The solid vertical red line in the right panel marks the Landau pole of the QED coupling at $\mu_L \approx 9.5$ GeV, and the dashed vertical red line the critical scale defined by~\eqref{eq:mucrit} at $\mu_c \approx 2.8$ GeV. Lastly, for every function defined at scales below $\mu_c$, the exponent $b_\mu$ remains below the red curve. In particular, for every such function there is a maximum scale of well-defined evolution before it is pushed to the critical region $b_\mu = -1$ (shaded gray) where the RG equation is ill-defined. }
	\label{fig:RGflow}
\end{figure}

We summarize this discussion in the left plot of Fig.~\ref{fig:RGflow} by showing the RG flow of the exponent $b_\mu = b - a(\mu,\mu_0)$ for a given $b$ at the reference scale $\mu_0$. In QCD-only, a power-like initial condition $\phi_M(u;\mu_0) \sim u^b$, evolved to higher scales $\mu > \mu_0$, results in an asymptotic expansion $\phi_M(u;\mu) \sim u^{b_\mu}$ as long as $b_\mu <1$, and $\phi_M(u;\mu) \sim u^1$ for $b_\mu \geq 1$. But once the linear endpoint behaviour is reached, the LCDA remains linear under further upward scale evolution, in agreement with the asymptotic form $\phi_M(u;\mu\to\infty)\to 6u\bar{u}$. Initial conditions with $b>1$ become automatically linear after one infinitesimal evolution step. Thus, linear endpoint behaviour can be viewed as a UV fixed point of the RG flow in QCD.  For electrically neutral mesons the QED anomalous dimension has the same form as in QCD with the replacement $\alpha_s C_F \to \alpha_s C_F+\alem Q_q^2$, and the above conclusions also hold (see also~\eqref{eq:neutralGegenbauers} and discussion thereafter). However, since QED becomes strongly coupled at (phenomenologically uninterestingly) large scales, the perturbative analysis breaks down and results based on 
the one-loop kernel are no longer reliable.

We will now follow the same steps to analyze to which extent the additional local logarithmic QED terms in~\eqref{eq:ERBL} alter this result for electrically charged mesons.
As we are only interested in the behaviour for small $u$, we normalize the LCDA to the point-like limit~\eqref{eq:Zell} to get rid of some $u$-independent local terms in the anomalous dimension. We define
\begin{align}
 \Phi_M(u;\mu) = Z_\ell(\mu) \hat{\Phi}_M(u;\mu) \,. 
\end{align}
The asymptotic RGE in Mellin-space for $\hat{\Phi}_M(u;\mu)$ is
\begin{align}
\label{eq:MellinRGE}
 \bigg[\frac{d}{d\ln\mu} \, + \, &\frac{\alpha_s C_F + \alpha_{\rm em} Q_{q_1} (Q_{q_2}-Q_M)}{\pi} \partial_\eta \bigg] \tilde{\hat{\Phi}}_M(\eta;\mu) \nonumber \\
 &= - \frac{\alpha_s C_F + \alpha_{\rm em} Q_{q_1} Q_{q_2}}{\pi} \left(H_\eta + H_{-\eta}-\frac32\right) \tilde{\hat{\Phi}}_M(\eta;\mu) \, ,
\end{align}
with the Mellin transform of the normalized LCDA $\tilde{\hat{\Phi}}_M(\eta;\mu)$. Adapting \cite{Liu:2020eqe}, the equation is solved by the ansatz
\begin{align}
\label{eq:mellinsolution}
\tilde{\hat{\Phi}}_M(\eta;\mu) &= \exp\left[2 \gamma_E a -3\check{a}/2\right]\; \frac{\Gamma(1-\eta) \Gamma(1+\eta+a)}{\Gamma(1+\eta)\Gamma(1-\eta-a)} \tilde{\hat{\Phi}}_M(\eta+a;\mu_0) \nonumber \\
&\times \exp\left\{ -\int_{\mu_0}^\mu \frac{d\mu'}{\mu'} \frac{\alpha_{\rm em}(\mu') Q_{q_1}Q_M}{\pi} \left( H_{\eta + a(\mu,\mu')} + H_{-\eta-a(\mu,\mu')} \right) \right\} \,,
\end{align}
The integrand of the exponent in the second line is proportional to the difference of charge factors on the left-hand side and right-hand side of~\eqref{eq:MellinRGE}, which arises as a consequence of the new $\ln u$ term in the evolution kernel. The generalized evolution variable reads
\begin{align}
\label{eq:adef}
 a \equiv a(\mu,\mu_0) = - \int_{\mu_0}^\mu \frac{d\tilde{\mu}}{\tilde{\mu}} \frac{\alpha_s(\tilde{\mu}) C_F + \alpha_{\rm em}(\tilde{\mu}) Q_{q_1} (Q_{q_2}-Q_M)}{\pi} \,.
\end{align}
In addition, we have to introduce a second variable which globally multiplies the solution and is thus irrelevant for the functional form in $u$: 
\begin{align}
\label{eq:acheckdef}
 \check{a} \equiv \check{a}(\mu,\mu_0) = - \int_{\mu_0}^\mu \frac{d\tilde{\mu}}{\tilde{\mu}} \frac{\alpha_s(\tilde{\mu}) C_F + \alpha_{\rm em}(\tilde{\mu}) Q_{q_1} Q_{q_2}}{\pi} \,.
\end{align}

It remains to understand the analytic properties of~\eqref{eq:mellinsolution} in the complex $\eta$-plane. The integral in the exponent in the second line can be solved analytically for some special cases that are instructive to study: i) for QED-only with one-loop running of $\alem(\mu)$ and ii) in the approximation of scale-independent gauge couplings in QCD$\times$QED. In both cases, we find a simple expression that is very similar to the QCD solution in~\eqref{eq:mellinsolutionQCD}:
\begin{align}
\label{eq:mellinsolutionpureQED}
\tilde{\hat{\Phi}}_M(\eta;\mu) &= \exp\left[(2 \gamma_E-3/2)\check{a}\right]  \left[\frac{\Gamma(1-\eta) \Gamma(1+\eta+a)}{\Gamma(1+\eta)\Gamma(1-\eta-a)}\right]^{1+p} \tilde{\hat{\Phi}}_M(\eta+a;\mu_0) \,,
\end{align}
but now with the combination of Gamma functions raised to the non-integer exponent \mbox{$p = \alem Q_{q_1} Q_M/(\alpha_s C_F + \alem Q_{q_1} (Q_{q_2}-Q_M))$}. 

Let us discuss the QED-only case i) first, for which $p$ reduces to $p =Q_M/(Q_{q_2}-Q_M) = -3/5$ for $Q_M = -1$. The important difference to QCD-only is that the evolution variable
\begin{align} \label{eq:aQED}
    a = Q_{q_1} (Q_{q_2}-Q_M) \frac{2}{\beta_0^{\rm QED}} \ln \frac{\alem(\mu)}{\alem(\mu_0)} +\mathcal{O}(\alem)
\end{align}
is always positive for an upward scale evolution to $\mu > \mu_0$. This is due to the quark charges and the sign of the one-loop QED $\beta$-function
\begin{equation}
\label{eq:betafctsQED}
    \frac{d \alem}{d \ln \mu} = - \frac{\alem^2}{2\pi}\beta_0^{\rm QED}\qquad\mbox{with}\qquad\beta_0^{\rm QED} = -\frac43 \left[ N_c (n_u Q_u^2 + n_d Q_d^2) +n_\ell Q_\ell^2 \right] <0 \, ,
\end{equation}
where $n_u (n_d)$ denotes the number of the active up (down) quark flavours and $n_\ell$ the active lepton flavours. The endpoint behaviour is then $\Phi_M(u;\mu) \sim u^{b_\mu}$ for $b < 1$, and $\Phi_M(u;\mu) \sim u^{1-a(\mu,\mu_0)}$ for $b \geq 1$. In particular, the exponent $b_\mu$ is driven towards smaller values, and, since $a$ increases without bound, will approach the point $b_\mu = -1$ for large enough $\mu$. This not only means that the LCDA becomes divergent at the endpoint, it also implies that RG evolution pushes the solution towards a functional form that is no longer compatible with the RGE itself, since for $b_\mu \leq -1$ the convolution integral in the evolution equation is no longer well-defined. Thus, QED evolution inevitably drives the solution outside the validity of its evolution kernel.

In the case ii) of fixed couplings in QCD$\times$QED, the evolution variable is given by 
\begin{align} \label{eq:afixed}
    a = - \frac{\alpha_s C_F + \alem Q_{q_1} (Q_{q_2} -Q_M)}{\pi} \ln \frac{\mu}{\mu_0} \; .
\end{align}
The QCD and QED terms enter with different signs due to the quark and meson charges. The behaviour of the solution \eqref{eq:mellinsolutionpureQED} now depends on whether the overall sign of \eqref{eq:afixed} is positive or negative. For $\alpha_s C_F + \alem Q_{q_1}(Q_{q_2} - Q_M)<0$, the evolution variable $a$ is positive and $b_\mu$ decreases with increasing $\mu$. The endpoint behaviour follows the same pattern as in the previously discussed QED-only case i). For the phenomenologically more relevant situation $\alpha_s C_F + \alem Q_{q_1}(Q_{q_2} - Q_M)>0$, the evolution variable $a$ is negative and $b_\mu$ increases with $\mu$. The situation is then similar to QCD-only, namely that the solution is pushed towards linear endpoint behaviour with $\Phi_M(u;\mu)\sim u^{b_\mu}$ for $b_\mu<1$, eventually reaching $b_\mu=1$ at some finite $\mu<\infty$. However, for $b_\mu=1$, one additional complication arises: the singularities of the gamma functions in~\eqref{eq:mellinsolutionpureQED} now have an associated branch cut. In particular, the left-most single pole from $\Gamma(1-\eta)$ has turned into a cut. After integration along this cut, we obtain
\begin{align}
\label{eq:lnp}
     \Phi_M(u;\mu) \sim u (-\ln u)^p   \,.
\end{align}
Hence, the linear endpoint behaviour is modified by a logarithmic term raised to the non-integer power $p$. In the limit $\alem \to 0$ the exponent $p$ vanishes and we recover the well-known linear endpoint behaviour in QCD. Expanding in $\alem$ to first order generates a term proportional to
$\alem u \ln\,(-\ln u)$.

Now, in the general case of QCD$\times$QED with one-loop running coupling the integrand of \eqref{eq:adef} naturally defines a critical scale by
\begin{align}
\label{eq:mucrit}
     \alpha_s(\mu_c) C_F + \alem(\mu_c) Q_{q_1} (Q_{q_2}-Q_M) = 0 \, .
\end{align}
At this scale, the exponent $p(\mu_c\mp 0)= \pm \infty$ becomes
singular and flips sign. This changes the analytic properties of the
Mellin-space RG solution. The resulting RG flow of $b_\mu$ is qualitatively shown in the right plot of Fig.~\ref{fig:RGflow}. Similar to case ii), it is useful to discuss the general solution \eqref{eq:mellinsolution} for $\mu_0>\mu_c$ and $\mu<\mu_c$ separately. For $\mu_0>\mu_c$, the left-hand side of \eqref{eq:mucrit} becomes negative, and the above discussion for fixed couplings with $\alpha_s C_F + \alem Q_{q_1} (Q_{q_2} -Q_M)<0$ applies, so that the endpoint behaviour is simply given by $\Phi_M(u;\mu) \sim u^{b_\mu}$ with decreasing $b_\mu<1$. Independent of $\mu_0$ and the functional form at this scale, the point where evolution becomes inconsistent is always reached at some finite scale $\mu_L > \mu_0$. At these extremely high scales, one enters the strong QED coupling regime and therefore this case is not relevant for realistic scenarios. However, in principle, in
theories with a different flavour content where the magnitude of $\beta_0^{\rm QED}$ is smaller, the evolution of $\Phi_M(u;\mu)$ can break down before entering the strong coupling regime. For $\mu<\mu_c$, the endpoint behaviour for increasing $b_\mu<1$ is again given by $\Phi_M(u;\mu) \sim u^{b_\mu}$. Based on the observation for fixed couplings, the particular point of interest is $b_\mu =1$. Remarkably, the result \eqref{eq:lnp} also holds in the general case of scale dependent gauge couplings in QCD$\times$QED. We prove this statement explicitly in Appendix~\ref{subsec:contourint}. An important consequence is that the inverse moments relevant for (QED-generalized) factorization theorems remain
integrable.

\begin{figure}[t]
    \centering
    \includegraphics[width=0.75\textwidth]{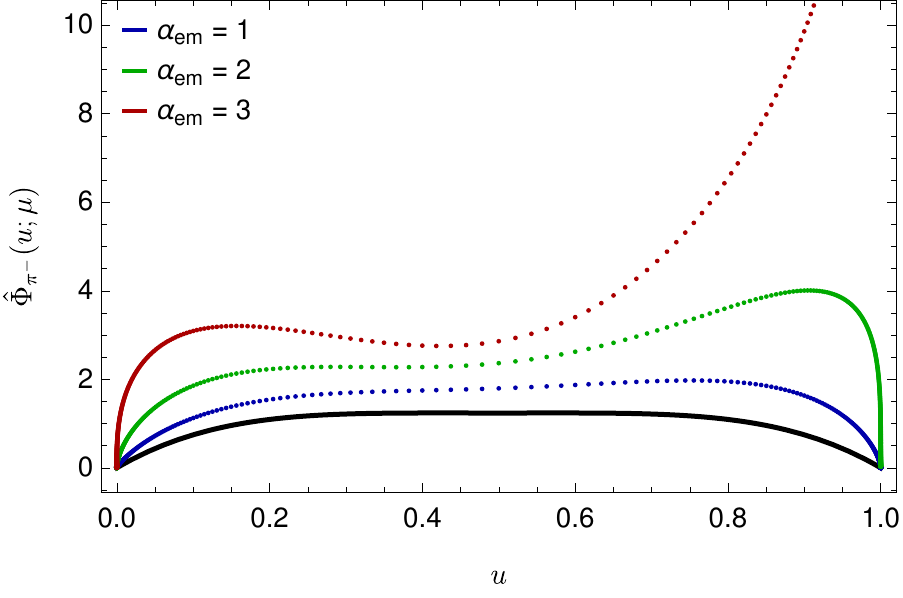}
    \caption{Numerical solution to the RGE for the $\pi^-$ LCDA by discretization in $u$ using $N = 1001$ logarithmically distributed points (see Section~\ref{subsec:pion:numestimates} for more details). We model the LCDA by the QCD-only initial condition  $\hat{\Phi}_{\pi^-}(u;\mu_{\rm lat}) = 6 u \bar{u} (1+a_2^\pi(\mu_{\rm lat}) C_2^{(3/2)}(2u-1))$ at the reference scale $\mu_{\rm lat} = 2$ GeV (black solid curve), using the lattice value for the Gegenbauer moment $a_2^\pi(\mu_{\rm lat})$ given in Table~\ref{tab:inputs}. The dotted curves show the function evolved to $\mu = 10$ GeV for three fictitiously large values of $\alem$. The QED coupling is fixed, while $\alpha_s(\mu)$ runs at one-loop. The following qualitative features induced by QED effects can be observed: First, the norm is no longer conserved. Second, the distribution becomes asymmetric and favours the up quark, which carries a larger fraction of the pion momentum due to its larger electromagnetic coupling. Third, the endpoint behaviour gets modified, and above a certain threshold the LCDA starts to diverge. 
    }
    \label{fig:endpointdiscrete}
\end{figure}

To qualitatively visualize the asymmetry of the QCD$\times$QED LCDA generated by evolution, we show in Fig.~\ref{fig:endpointdiscrete} a numerical solution of the RGE obtained from solving the integro-differential evolution equation \eqref{eq:pionRG} by discretization (see Section~\ref{subsec:pion:numestimates} for more details) for fictitious values of the electromagnetic coupling constant. 

\section{Gegenbauer coefficients and analytic $\mathcal{O}(\alem)$ solution}
\label{sec:geg}

In the real world, QED effects are expected to be small. In this section we provide an analytic solution to the QCD$\times$QED RGE that treats QED effects at first order while summing $\ln \mu/\mu_0$ to all orders, i.e. a solution accurate to $\alem^k \alpha_s^n \ln^{n+k} \mu/\mu_0$ with $k=0,1$. For this purpose, we expand the LCDA in Gegenbauer polynomials as usual such that the integro-differential evolution equation~\eqref{eq:pionRG} becomes an infinite dimensional system of ordinary differential equations of Gegenbauer coefficients $a_n^M(\mu)$. To separate universal from structure-dependent QED effects we again normalize to the point-like limit and consider 
\begin{equation}
\label{eq:QCDgegenansatz}
    \hat{\Phi}_M(u;\mu) = 6u \bar{u} \sum_{n=0}^\infty a^M_n(\mu) C_n^{\left( 3/2 \right)}(2u-1) \; ,
\end{equation}
where $C_n^{(3/2)}(2u-1)$ are Gegenbauer polynomials and $a_n^M(\mu)$ the $n$-th Gegenbauer coefficients for meson $M$. 
The RGE for the coefficients $a_n^M(\mu)$ reads
\begin{equation}
\label{eq:QEDmoments}
    \frac{d}{d \ln \mu} a^M_n(\mu) = - \frac{\alpha_s(\mu) C_F + \alem(\mu) Q_{q_1} Q_{q_2}}{2\pi} \gamma_n a^M_n(\mu) - \frac{\alem(\mu)}{\pi} Q_M \sum_{m=0}^\infty f_{nm} a^M_m(\mu) \; .
\end{equation}
For neutral mesons, $Q_{q_1}=Q_{q_2}=Q_q$ and $Q_M = 0$, the scale evolution of the $a_n^M(\mu)$ is diagonal and solved by
\begin{align}
\label{eq:neutralGegenbauers}
    a_n(\mu) = \left( \frac{\alpha_s(\mu)}{\alpha_s(\mu_0)}  \right)^{C_F \gamma_n/\beta^{\rm QCD}_0} \left( \frac{\alem(\mu)}{\alem(\mu_0)}  \right)^{Q_q^2 \gamma_n /\beta^{\rm QED}_0} a_n(\mu_0) 
\end{align}
in the leading-logarithmic (LL) approximation. The $n$-dependent anomalous dimension reads~\cite{Lepage:1979zb}
\begin{align}
    \gamma_n = 1-\frac{2}{(n+1)(n+2)} +4\sum_{m=2}^{n+1} \frac{1}{m} = 4 H_{n+1} -3 -\frac{2}{(n+1)(n+2)}
\end{align}
and grows logarithmically for large $n$, $\gamma_n \approx 4 \ln(n)$. In QCD, the evolution to larger values of $\mu$ suppresses higher Gegenbauer coefficients since $\alpha_s(\mu)<\alpha_s(\mu_0)$ and the exponent $C_F \gamma_n/\beta_0^{\rm QCD}$ in \eqref{eq:neutralGegenbauers} is positive and growing with $n$. The QED factor in \eqref{eq:neutralGegenbauers} additionally suppresses the coefficients as $\alem(\mu)>\alem(\mu_0)$ and $Q_q^2 \gamma_n/\beta_0^{\rm QED}$ is negative. This justifies the truncation of the Gegenbauer series at some fixed value $n=n_0$ for neutral mesons even after QED corrections are included. In QCD only, one typically chooses $n_0=2$, up to which data from lattice QCD is available. Note that $\gamma_0 = 0$, so $a_0$ is not renormalized for neutral mesons. In fact, the normalization condition $\int_0^1 du\, \Phi_M(u;\mu) = 1$ implies $a_0 = 1$.

For charged mesons we observe two important differences. First, the QED contribution enters the first term in~\eqref{eq:QEDmoments} with a different sign.
However, at physically relevant scales, $\alem(\mu)$ can be viewed as a small perturbation compared to $\alpha_s(\mu)$, justifying a truncation also in that case.\footnote{If one evolves to (phenomenologically uninteresting) high scales, $\alem(\mu)$ eventually dominates over $\alpha_s(\mu)$. The overall sign-flip of the diagonal terms in the anomalous dimension then leads to the opposite conclusion: higher Gegenbauer coefficients become more and more important.} The solution to~\eqref{eq:QEDmoments} can then be obtained by numerically solving the resulting finite-dimensional system of first-order differential equations.  Second, for $Q_M \neq0$ the local $\ln u$ and $\ln \bar{u}$ terms in the anomalous dimensions cause a mixing of Gegenbauer coefficients under RG evolution, with the infinite-dimensional mixing matrix $f_{nm}$ given by
\begin{align} \label{eq:QEDmixingcoeff} 
    f_{nm} &= \frac{4(2n+3)}{(n+2)(n+1)} \int_0^1 du~(Q_{q_1} \ln u - Q_{q_2} \ln \bar{u}  )~u\bar{u}~C^{\left( 3/2 \right)}_n(2u-1) C^{\left( 3/2 \right)}_m(2u-1) \nonumber \\ 
    &=  \left( Q_{q_2} -(-1)^{n+m} Q_{q_1} \right) \times \begin{cases} 
\displaystyle 
\frac{(2n+3)}{(n-m)(n+m+3)} \times \frac{(m+1)(m+2)}{(n+1)(n+2)}\qquad & n > m\\[0.5cm] 
\displaystyle \frac{(2n+3)}{(m-n)(n+m+3)} & n < m\\[0.5cm] 
\displaystyle \frac{1}{2n+3} + H_{n+1/2} - H_{n+2} + \ln 4 & n=m  \,.
    \end{cases}
\end{align}
Contrary to the triangular structure of the anomalous dimension matrices in QCD, the mixing matrix $f_{nm}$ has no particular structure. This means that in QED the Gegenbauer coefficients $a_n$ can also mix into lower coefficients $a_m$ with $m<n$. Since $f_{0m} \neq 0$, there is even a mixing of all Gegenbauer coefficients into the zeroth coefficient $a_0^M(\mu)$, such that the standard normalization condition of the LCDA is no longer valid for charged light mesons due to scale evolution, i.e. $\int_0^1 du \, \Phi_M(u;\mu) \neq 1$. The diagonal terms $f_{nn}$ approach the constant $Q_M \ln 4$ and are thus negligible compared to the logarithmically growing first term in~\eqref{eq:QEDmoments} for large $n$. In all other cases, $f_{nm}$ falls off for large $n$ or $m$, indicating that the mixing of Gegenbauer coefficients with largely different $n$ is strongly suppressed (even for large values of $\alem(\mu)$). For example, for $n \gg m$, the mixing matrix drops like $f_{nm} \sim 1/n^3$, and for $m \gg n$ one has $f_{nm} \sim 1/m^2$. Also for $n$ and $m$ large but of the 
same order the off-diagonal terms fall off like $ 1/n$.

For scales relevant to hard exclusive processes it is sufficient to consider $\alem(\mu)$ as a small perturbation.
In the following we provide an analytic solution to the RGE for the Gegenbauer coefficients to first order in $\alem$. In other words, we sum large logarithms $L$ to all orders in QCD and count $\alpha_s \times L \sim \mathcal{O}(1)$, but retain the fixed-order expansion in $\alem$, since $\alem \times L \ll 1$ is still small. This corresponds to the summation of the leading logarithms $\alem^k \alpha_s^n \ln^{n+k}\mu/\mu_0$ with $k=0,1$. However, we always 
retain the resummed expression for the large QED double-logarithms in~\eqref{eq:uell} associated with the point-like limit. To this end, we expand the Gegenbauer coefficients  as
\begin{equation}
    a_n(\mu) = a_n^{\rm QCD}(\mu) + \frac{\alem(\mu)}{\pi} a_n^{(1)}(\mu) + \mathcal{O}(\alem^2) \ .
\label{eq:QEDanexp}
\end{equation}
The QCD moments $a^{\rm QCD}_n(\mu) \sim \mathcal{O}(1)$ count as order one. If one expands the initial condition at $\mu_0$ in the 
above form, $a^{(1)}_n(\mu_0) \sim \mathcal{O}(1)$, and 
the QED initial condition is a correction beyond the LL order. 
However, as will be seen from the solution below, 
evolution generates a QED correction  $a_n^{(1)}(\mu) \sim  
\mathcal{O}(\ln (\mu/\mu_0), \alpha_s(\mu_0)/\alpha_s(\mu))$, 
which identifies the second term on the right-hand side of 
\eqref{eq:QEDanexp} as the 
LL term linear in the QED coupling. The RGE in~\eqref{eq:QEDmoments} evidently reduces to the standard QCD evolution equation
\begin{align}
\label{eq:momentDGL0}
    \frac{d}{d\ln\mu}a_n^{\rm QCD}(\mu) &= - \frac{\alpha_s(\mu) C_F}{2\pi} \gamma_n a_n^{\rm QCD}(\mu)
\end{align}
at zeroth order in $\mathcal{O}(\alem)$, and the inhomegenous 
equation 
\begin{align}
\label{eq:momentDGL1}
    \frac{d}{d\ln\mu}a_n^{(1)}(\mu) &= - \frac{\alpha_s(\mu) C_F}{2\pi} \gamma_n a_n^{(1)}(\mu) - I_n(\mu)
\end{align}
for the first-order coefficients in the electromagnetic coupling $\alem$. The inhomogeneous term reads 
\begin{align}
    I_n(\mu) = \frac12 Q_{q_1} Q_{q_2} \gamma_n a_n^{\rm QCD}(\mu) + Q_M \sum_{m=0}^\infty f_{nm} a_m^{\rm QCD}(\mu)\,, 
\end{align}
and depends on \emph{all} QCD Gegenbauer coefficients at the scale $\mu$. The index $M$ indicating the light meson has been dropped on the $a_n(\mu)$ for better readability. The solution 
to \eqref{eq:momentDGL0} recovers the standard LL QCD expression 
\begin{align}
\label{eq:LLGegenbauers}
    a_n^{\rm QCD}(\mu) = \left(\frac{\alpha_s(\mu)}{\alpha_s(\mu_0)}\right)^{C_F \gamma_n/\beta^{\rm QCD}_0} a_n^{\rm QCD} (\mu_0) \,.
\end{align}
The desired LL solution for the linear QED correction reads
\begin{eqnarray}
\label{eq:QEDfirstmoment}
   a_n^{(1)}(\mu) &=&  \left(\frac{\alpha_s(\mu)}{\alpha_s(\mu_0)}\right)^{C_F \gamma_n/\beta^{\rm QCD}_0}a_n^{(1)}(\mu_0) -\frac12 Q_{q_1} Q_{q_2} \gamma_n a_n^{\rm QCD}(\mu) \ln \frac{\mu}{\mu_0} \nonumber \\ 
&&\hspace*{-1cm}-  \,Q_M \sum_{m=0}^\infty \frac{2\pi  f_{nm} }{\beta_0^{\rm QCD}+(\gamma_n-\gamma_m)C_F} \left\{ \frac{a_m^{\rm QCD}(\mu)}{\alpha_s(\mu)} - \frac{a_m^{\rm QCD}(\mu_0)}{\alpha_s(\mu_0)}  \left( \frac{\alpha_s(\mu)}{\alpha_s(\mu_0)}  \right)^{C_F \gamma_n/\beta_0^{\rm QCD}} \right\}.\qquad
\end{eqnarray}
As mentioned above, the QED initial condition $a_n^{(1)}(\mu_0)$ 
is technically beyond the LL accuracy. Also, 
if the input values at the scale $\mu_0$ are given by a certain 
model or (future) lattice calculations in
 QCD$\times$QED, it does not seem very natural to expand the 
initial condition itself in $\alem$. In both cases, one 
sets $a_n^{(1)}(\mu_0) \to 0$ in \eqref{eq:QEDfirstmoment}, 
as we do in the numerical analysis in the next section.

\section{Numerical estimates}
\label{subsec:pion:numestimates}
  
In this section we provide more details on the discretization of the RGE used to obtain Fig.~\ref{fig:endpointdiscrete} and give numerical estimates of the QED corrections to the Gegenbauer coefficients and the inverse moments of 
\begin{align}
\Phi_M(u;\mu) =  Z_\ell(\mu) \,6u \bar{u} \sum_{n=0}^\infty a^M_n(\mu) C_n^{\left( 3/2 \right)}(2u-1)  \; ,
\end{align} which are relevant to (QED-generalized) factorization theorems for hard exclusive processes. We recall that the LCDA (but not its evolution) is IR divergent for electrically charged mesons, as it is part of the non-radiative amplitude and only accounts for virtual QED contributions. The initial conditions employed below should be interpreted as a model for the properly IR-subtracted, and thus scheme-dependent, LCDA at the scale $\mu_0$. Under this assumption, the UV scale evolution can be studied for which we give numerical estimates. 
We use the input values given in Table~\ref{tab:inputs}. For the gauge couplings, we decouple the bottom (charm) quark in $\alpha_s(\mu)$ at its pole mass $\mu = m_b$ ($\mu = m_c$). For the electromagnetic coupling $\alem(\mu)$, we also include the threshold at $m_\tau$. 

\begin{table}[t]
	\begin{center}
		\begin{tabularx}{0.98\textwidth}{|C C C|}
	\hline 
	\multicolumn{3}{|c|}{Coupling constants and $Z$ boson mass}\\
		\hline 
  $\alem(m_Z)=1/127.96$  & $\alpha_s(m_Z) = 0.1181$ & $m_Z=91.1876$ GeV
  \\ \hline
		\end{tabularx}
				\vskip 1pt
		\begin{tabularx}{0.98\textwidth}{|C C C|}
	\hline 
	\multicolumn{3}{|c|}{Quark and lepton masses}\\
		\hline 
  $m_b=4.78$ GeV  & $m_c = 1.67$ GeV & $m_\tau=1.78$ GeV
  \\ \hline
		\end{tabularx}
		\vskip 1pt
		\begin{tabularx}{0.98\textwidth}{|C C C|}
\hline	\multicolumn{3}{|c|}{Gegenbauer coefficients at $\mu_{\rm lat}=2$ GeV}\\
		\hline 	
$a_0^{\pi}=1$ & $a_1^{\pi}=0$ & $a_2^{\pi} = 0.116^{+19}_{-20}$ \\ \hline
$a_0^{K}=1$ & $a_1^{K}=0.0525^{+31}_{-33}$ & $a_2^{K} = 0.106^{+15}_{-16}$ \\ \hline
\end{tabularx}
		\vskip 1pt
		\begin{tabularx}{0.98\textwidth}{|C C C|}
\hline	\multicolumn{3}{|c|}{Gegenbauer coefficients at $\mu_0=1$ GeV}\\
		\hline 	
$a_0^{\pi}=1$ & $a_1^{\pi}=0$ & $a_2^{\pi} = 0.140^{+23}_{-24}$ \\ \hline
$a_0^{K}=1$ & $a_1^{K}=0.0593^{+35}_{-37}$ & $a_2^{K} = 0.128^{+18}_{-19}$ \\ \hline
\end{tabularx}
\caption{Numerical inputs. The Gegenbauer coefficients at $\mu_{\rm lat} =2\,$GeV are the lattice QCD results from~\cite{Bali:2019dqc}. The quark masses are to be understood as two-loop pole masses.}
\label{tab:inputs}
	\end{center}
\end{table}

To discretize the evolution equation \eqref{eq:pionRG}, we divide the interval $u\in [0,1]$ into $N-1$ sub-intervals by distributing $N$ points $u_i$, with $i=1, \dots, N$, between $0$ and $1$. For large $N$ the integral in $v$ can be approximated by a Riemann sum and the integro-differential equation turns into a coupled system of $N$ ordinary first-order differential equations which we solve numerically. To increase the accuracy, we use the trapezoidal rule, whose error is roughly proportional to the third power  $(\Delta u)^3$ of the 
difference $\Delta u$ of two points. As the integration kernel is divergent at the endpoints $u=0$ and $u=1$, we shrink the interval $u\in [0,1]$ to $u\in[\eps,1-\eps]$, with $\eps \ll 1$. Especially when studying the endpoint behaviour, it is important to choose $\eps$ sufficiently small.
As a default value we use $\eps=10^{-10}$. To increase the accuracy it is useful to distribute points non-uniformly in such a way that the density of points is logarithmically enhanced towards the endpoints. Using, for example, $N = 1001$ points, we reproduce the known analytic LL solution in QCD with an error of less than $0.04$\% for the initial condition and scale choice described in the caption of Fig.~\ref{fig:endpointdiscrete}. We emphasize that this error analysis serves as a cross-check of the implementation, but provides only a rough estimate of the expected accuracy in QED, as contributions from the endpoints can be enhanced for large, but unphysical values of $\alem$.

To estimate the size of QED contributions, we compare the scale evolution of the Gegenbauer coefficients in pure QCD at the LL, NLL, and NNLL order with the evolution in QCD$\times$QED using the first-order QED solution~\eqref{eq:QEDfirstmoment}.\footnote{In the QCD NNLL evolution, we neglect the unknown matching relation of the Gegenbauer coefficients at the flavour threshold $m_b$, which is expected to give a small correction to the NNLL results.} For the NLL evolution, we use the NLO anomalous dimension matrix and the two-loop running coupling $\alpha_s$, whereas the NNLO anomalous dimension and three-loop running is used for the NNLL results. In QCD, we use the input values for the Gegenbauer coefficients at $\mu_{\rm lat}= 2$ GeV obtained from lattice QCD \cite{Bali:2019dqc} given in Table~\ref{tab:inputs}, and evolve to a higher scale $\mu$. In QCD$\times$QED, we first run down to the natural scale $\mu_0 = 1$ GeV at LL in pure QCD, resulting in the values given in Table~\ref{tab:inputs} and then evolve to $\mu$ including QED. We give numerical results for two different scales: $\mu = 5.3 \,\text{GeV} \approx m_B$ and $\mu = 80.4 \,\text{GeV} \approx M_W$. The first is relevant for exclusive $B$ decays, the latter would be relevant e.g. for the very rare $W^- \to \pi^- \gamma$ decay~\cite{Grossmann:2015lea}. 

In QED all Gegenbauer coefficients at the low scale $\mu_0$ with $n>2$ mix into $a_{0,1,2}(\mu)$. However, as currently the Gegenbauer coefficients with $n>2$ are unknown and since they are expected to be small, we set them to zero at the initial scale $\mu_0$. In addition, these higher Gegenbauer coefficients will be generated through scale evolution in QED, and also in QCD beyond the LL accuracy. We include this mixing when calculating the inverse moments, as discussed below. At the moment, also the QED corrections to the Gegenbauer coefficients at the reference scale $\mu_0=1$ GeV are unknown. Nevertheless, as discussed above, the dominant logarithmically enhanced (LL) contributions are captured by evolution.

For a negatively charged pion $\pi^-=(d\bar{u})$, we obtain at $\mu = 5.3$ GeV the values \begin{align}
    a_0^{\pi^-} &= 1 + 1.44 \cdot \frac{\alem(\mu)}{\pi} = 1\big\vert_{\rm QCD} + 0.0035\big\vert_{\rm QED} \nn \,, \\ 
    a_1^{\pi^-} &= 0 + 0.25 \cdot \frac{\alem(\mu)}{\pi} = 0.0006\big\vert_{\rm QED} \,,\\
    a_2^{\pi^-} &= 0.0951\big\vert_{\rm LL} - 0.0084\big\vert_{\rm NLL} + 0.0001\big\vert_{\rm NNLL} + 0.42 \cdot \frac{\alem(\mu)}{\pi} = 0.0867\big\vert_{\rm QCD} + 0.0010\big\vert_{\rm QED}   \nn \,,
\end{align}
and for the kaon $K^-=(s\bar{u})$
\begin{align}
    a_0^{K^-} &= 1 + 1.46\cdot \frac{\alem(\mu)}{\pi} = 1\big\vert_{\rm QCD} + 0.0035\big\vert_{\rm QED} \,, \\ 
    a_1^{K^-} &= 0.0462\big\vert_{\rm LL} - 0.0023\big\vert_{\rm NLL} + 0.0001\big\vert_{\rm NNLL} + 0.35 \cdot \frac{\alem(\mu)}{\pi} = 0.0441\big\vert_{\rm QCD} + 0.0009\big\vert_{\rm QED} \nn  \,,\\
    a_2^{K^-} &= 0.0869\big\vert_{\rm LL} - 0.0078\big\vert_{\rm NLL} - 0.0000\big\vert_{\rm NNLL} + 0.41 \cdot \frac{\alem(\mu)}{\pi} = 0.0791\big\vert_{\rm QCD} + 0.0010\big\vert_{\rm QED} \,. \nn 
\end{align}
At $\mu = 80.4$ GeV
\begin{align}
    a_0^{\pi^-} &= 1 + 3.78 \cdot \frac{\alem(\mu)}{\pi} = 1\big\vert_{\rm QCD} + 0.0094\big\vert_{\rm QED} \nn \,, \\ 
    a_1^{\pi^-} &= 0 + 0.59 \cdot \frac{\alem(\mu)}{\pi} = 0.0015\big\vert_{\rm QED}\,, \\
    a_2^{\pi^-} &= 0.0657\big\vert_{\rm LL} - 0.0098\big\vert_{\rm NLL} + 0.0002\big\vert_{\rm NNLL} + 0.84 \cdot \frac{\alem(\mu)}{\pi} = 0.0561\big\vert_{\rm QCD} + 0.0021\big\vert_{\rm QED}  \, ,
\nn \end{align}
and
\begin{align}
    a_0^{K^-} &= 1 + 3.82 \cdot \frac{\alem(\mu)}{\pi} = 1 + 0.0095\big\vert_{\rm QED} \,,\\ 
    a_1^{K^-} &= 0.0365\big\vert_{\rm LL} - 0.0030\big\vert_{\rm NLL} + 0.0002\big\vert_{\rm NNLL} + 0.80 \cdot \frac{\alem(\mu)}{\pi} = 0.0336\big\vert_{\rm QCD} + 0.0020\big\vert_{\rm QED} \nn \,,\\
    a_2^{K^-} &= 0.0601\big\vert_{\rm LL} - 0.0091\big\vert_{\rm NLL} + 0.0002\big\vert_{\rm NNLL} + 0.83 \cdot \frac{\alem(\mu)}{\pi} = 0.0511\big\vert_{\rm QCD} +0.0021\big\vert_{\rm QED} \,.
\nn \end{align}
We refrain from quoting uncertainties as the current uncertainty on the lattice input values at $\mu_{\rm lat} = 2$ GeV is $\mathcal{O}(15\%)$, which is larger than the QCD NLL contribution in most cases. Our aim is to compare the relative size of QED effects to higher-order evolution in QCD. We find that the QED effects are almost an order of magnitude larger than the NNLL evolution, although consistently below the NLL contribution. For example, the relative contribution of the QED effects in $a_2^{\pi^-}$ rises from roughly $1\%$ at $5.3$ GeV to almost $4\%$ at $80.4$ GeV. Compared to the NLL contribution it increases from roughly $12\%$ to $21\%$. When evolving to higher scales, the QED effects become relatively more important, because the absolute value of the QCD coupling decreases, whereas simultaneously the QED value increases slightly. In fact, in QCD-only, the value of all Gegenbauer coefficients will tend to zero in this upward scale evolution and the LCDA approaches the asymptotic form $\phi_M(u;\mu\to\infty) = 6 u \bar{u}$. This does not hold for QED. In addition, QED modifies coefficients that are fixed to all orders in QCD by isospin symmetry (e.g. $a_1^{\pi^-} = 0$) or by the normalization of the LCDA ($a_0^M = 1$).   

Besides the Gegenbauer moments themselves, the inverse moments of the LCDA are important, as they often appears at leading order in factorization theorems for observables. They are defined by 
\begin{align}
\label{eq:ubarinverse}
     \left\langle \bar{u}^{-1} \right\rangle_{M^-}(\mu)  &= \int_0^1 \frac{du}{1-u} \Phi_{M^-}(u;\mu) = 3 Z_\ell(\mu) \sum_{n=0}^\infty a_n^{M^-}(\mu)\; , \\
\label{eq:uinverse}
     \left\langle u^{-1} \right\rangle_{M^-}(\mu) &= \int_0^1 \frac{du}{u} \Phi_{M^-}(u;\mu) = 3 Z_\ell(\mu) \sum_{n=0}^\infty (-1)^n a_n^{M^-}(\mu) \; .
\end{align}
and represented by an infinite sum over Gegenbauer coefficients. 
For the $\pi^-$, at the scales $\mu=5.3$ and $80.4$ GeV, we find
\begin{align}
     \left\langle \bar{u}^{-1} \right\rangle_{\pi^-}(5.3 \, \text{GeV}) &= 0.9997\big\vert^{\rm QED}_{\rm point\,charge} ( 3.285^{+0.05}_{-0.05}\big\vert_{\rm LL} - 0.020\big\vert_{\rm NLL}  + 0.017\big\vert^{\rm QED}_{\rm partonic}) \; , \nn \\
     \left\langle u^{-1} \right\rangle_{\pi^-}(5.3 \, \text{GeV}) &= 0.9997\big\vert^{\rm QED}_{\rm point\,charge} (3.285^{+0.05}_{-0.05}\big\vert_{\rm LL} - 0.020\big\vert_{\rm NLL}  + 0.012\big\vert^{\rm QED}_{\rm partonic})  \;, 
\end{align}
and
\begin{align}
     \left\langle \bar{u}^{-1} \right\rangle_{\pi^-}(80.4 \, \text{GeV}) &= 0.985\big\vert^{\rm QED}_{\rm point\,charge} ( 3.197^{+0.03}_{-0.03}\big\vert_{\rm LL} - 0.022\big\vert_{\rm NLL}  + 0.042\big\vert^{\rm QED}_{\rm partonic}) \; , \nn \\
     \left\langle u^{-1} \right\rangle_{\pi^-}(80.4 \, \text{GeV}) &= 0.985\big\vert^{\rm QED}_{\rm point\,charge} ( 3.197^{+0.03}_{-0.03}\big\vert_{\rm LL} - 0.022\big\vert_{\rm NLL}  + 0.031\big\vert^{\rm QED}_{\rm partonic})  \;.
\end{align}
For the charged kaon $K^-$, we find
     \begin{align}
     \left\langle \bar{u}^{-1} \right\rangle_{K^-}(5.3 \, \text{GeV}) &= 0.9997\big\vert^{\rm QED}_{\rm point\,charge} ( 3.399^{+0.05}_{-0.05}\big\vert_{\rm LL} - 0.026\big\vert_{\rm NLL}  + 0.018\big\vert^{\rm QED}_{\rm partonic}) \; , \nn \\
     \left\langle u^{-1} \right\rangle_{K^-}(5.3 \, \text{GeV}) &= 0.9997\big\vert^{\rm QED}_{\rm point\,charge} (3.122^{+0.03}_{-0.03}\big\vert_{\rm LL} - 0.011\big\vert_{\rm NLL}  + 0.011\big\vert^{\rm QED}_{\rm partonic})  \;,
\end{align}
and
\begin{align}
     \left\langle \bar{u}^{-1} \right\rangle_{K^-}(80.4 \, \text{GeV}) &= 0.985\big\vert^{\rm QED}_{\rm point\,charge} ( 3.290^{+0.03}_{-0.03}\big\vert_{\rm LL} - 0.029\big\vert_{\rm NLL}  + 0.044\big\vert^{\rm QED}_{\rm partonic}) \; , \nn \\
     \left\langle u^{-1} \right\rangle_{K^-}(80.4 \, \text{GeV}) &= 0.985\big\vert^{\rm QED}_{\rm point\,charge} (3.071^{+0.02}_{-0.02}\big\vert_{\rm LL} - 0.010\big\vert_{\rm NLL}  + 0.029\big\vert^{\rm QED}_{\rm partonic})  \;.
\end{align}
Here we separated the effect arising from the point-like limit 
contained in $Z_\ell(\mu)$, denoted by ``point charge'', from the structure-dependent contributions in the Gegenbauer coefficients, denoted by  ``partonic''. Again, we keep the resummed form for $Z_\ell(\mu)$ in~\eqref{eq:uell} but use the fixed-order $\mathcal{O}(\alem)$ solution~\eqref{eq:QEDfirstmoment} for the Gegenbauer coefficients.\footnote{Since we use QCD-only input at the low scale $\mu_0$, we set $Z_\ell(\mu_0) = 1$.} We set the energy $E$ that enters the evolution of $Z_\ell(\mu)$ to $E = \mu/2$, which is the hard scale in two-body decays. We note that the inverse moments in~\eqref{eq:ubarinverse} and~\eqref{eq:uinverse} depend on an infinite sum of Gegenbauer coefficients, which were set to zero at the initial scale $\mu_0$ for $n>2$. Scale evolution, both in QED and QCD at NLL, will then generate these higher Gegenbauer moments, which we include up to $n_{\rm max}=100$. We find that the sum for QCD NLL converges rather slowly. However, based on a naive convergence analysis, increasing the number of Gegenbauer coefficients may change the last digit of our results by at most one. 
The convergence in QED is much better and in almost all cases it was sufficient to truncate at a maximum value $n_{\rm max} = 10$ to obtain the quoted result. We note that we do not include an additional uncertainty from neglecting unknown Gegenbauer coefficients with $n>2$ at the reference scale $\mu_{\rm lat} = 2$ GeV. In this sense, our numerical analysis should be again interpreted as an estimate of the relative effects of higher order evolution in QCD versus QED effects, given a model at the low scale.

For the inverse moments the relative size of the various effects is different from those observed for the individual Gegenbauer coefficients. We first note that both QCD NLL as well as QED effects are typically of $\mathcal{O}(1\%)$ and of the same size as the uncertainty from the lattice values given in Table~\ref{tab:inputs}. 
For comparison, we therefore provide the uncertainty from the lattice Gegenbauer moment input for the QCD LL results, but do not give the corresponding errors on the QCD NLL and QED as well as NNLL contributions here, which are only a subleading correction. The structure-dependent QED effects become larger than the NLL QCD correction from evolution above scales of order $\mathcal{O}(10$ GeV$)$. At the high scale $80.4$ GeV it also exceeds the uncertainty from the current lattice determination of the input values. This can be understood from the fact that, unlike QCD, QED evolution couples all $a_n^M$ to the zeroth Gegenbauer coefficient $a_0^M$,~i.e. the norm of the LCDA, which is an order of magnitude larger than the first and second Gegenbauer coefficients. For example, at $\mu = 80.4$ GeV for the $\pi^-$, QED corrections cause an $\mathcal{O}(1\%)$ effect on $a_0^M = 1$, whereas higher-order QCD evolution is a $15\%$ effect on the LL value $a_2^{\pi^-} = 0.0657$, which enhances the relative size of QED effects. In comparison, the isospin breaking effects arising from QED due to a difference $\left\langle \bar{u}^{-1} - u^{-1} \right\rangle_{\pi^-}$ are at the few permille to 1\% level.

Lastly, we note that the structure-dependent QED corrections enhance the value for the inverse moments, whereas the contribution from the point charge acts in the opposite direction. The separation of the two effects is nervertheless useful, since the point-charge contribution is typically factored out and by definition not considered as part of the non-radiative amplitude.


\section{Conclusion}
\label{sec:conclusion}

In this paper, we studied the evolution of the leading-twist light-cone distribution amplitude (LCDA) for light mesons in QCD$\times$QED. This QED-generalized LCDAs was  introduced as part of the generalization of the QCD factorization formula for non-leptonic decays to QCD$\times$QED \cite{Beneke:2021jhp, Beneke:2020vnb}. We solved the (one-loop) RGE numerically and provided analytical expressions for the RGE at $\mathcal{O}(\alem)$ which resum the large logarithms in QCD on top of the fixed-order expansion in $\alem$. 

For electrically neutral mesons, the RGE kernel is simply the 
QCD ERBL evolution kernel~\cite{Lepage:1979zb, Lepage:1980fj,Efremov:1979qk} with the modified coupling  $\alpha_s C_F \to \alpha_s C_F+\alem Q_q^2$, where $Q_q$ denotes the electric charge of the 
quark in the meson and the evolution is similar to the 
case of QCD. 
For electrically charged mesons, however, 
the solution to the RGE exhibits interesting, qualitatively 
novel features, which ultimately arise from the coupling 
of soft photons to the net charge of the meson:
\begin{itemize}
\item QED effects change the endpoint behavior of the LCDA as 
shown in Fig.~\ref{fig:endpointdiscrete}. The linear 
vanishing of the LCDA near the endpoints $u=0,1$ is no longer 
the UV fixed point of the evolution.
\item  Due to new local logarithmic terms in the RGE, the evolution  of the Gegenbauer coefficients is no longer of triangular form in Gegenbauer moment space. Opposed to QCD, it induces mixing of higher Gegenbauer coefficients into lower ones. Moreover, the normalization condition $a_0^M=1$ 
is not stable under RGE evolution and can no longer be 
imposed. This is a manifestation of the 
scale dependence of the analogue of the decay constant 
for a charged pion in QCD$\times$QED (defined by $t=0$ in 
\eqref{eq:M2LCDA}). 
\item  The QED evolution violates isospin symmetry and the LCDA of charged $\pi$ mesons becomes antisymmetric. The QED-generalized LCDA favours larger momentum of the up-type quark due to the larger modulus of its charge.
\item The QED-generalized LCDA is not boost-invariant and 
depends on the energy of the meson, measured in a soft 
reference frame, determined by the process under 
consideration. The energy dependence appears in universal 
double-logarithmic terms, which can be factored out and equal the contribution of an electrically charged point 
particle.
\end{itemize}

In addition to the investigation of the RGE solution, 
we presented numerical estimates of the QED corrections from 
evolution for the Gegenbauer coefficients and the inverse moments, relevant for hard exclusive processes, at the two scales $\mu = 5.3$ and $\mu=80.4$ GeV. For the Gegenbauer coefficients, we found that the QED effects are almost one order magnitude larger than NNLL evolution. For the inverse moments, both the QED effects and the QCD NLL are at the percent level, and of similar size as the uncertainties from the lattice input values. For these moments, we separated the QED effects in structure-dependent and point-charge terms, which is useful in light of the QCD$\times$QED factorization formulas, where the QED-generalized LCDA naturally appears 
\cite{Beneke:2020vnb,Beneke:2021jhp}.

\subsubsection*{Acknowledgements}
We are grateful to Yao Ji for providing us with code on the three-loop evolution of Gegenbauer coefficients, based on~\cite{Strohmaier:2018tjo}. This research was supported in part by the Deutsche
Forschungsgemeinschaft (DFG, German Research Foundation) through
the Sino-German Collaborative Research Center TRR110 “Symmetries
and the Emergence of Structure in QCD”
(DFG Project-ID 196253076, NSFC Grant No. 12070131001, - TRR 110). J.-N. T. would like to thank
the “Studienstiftung des deutschen Volkes” for a scholarship.

\appendix

\section{Soft rearrangement}
\label{app:sec:notation}
In this appendix, we calculate the subtraction factors $R_c$ and $R_{\bar{c}}$ in \eqref{eq:softsubtraction},
\begin{equation}
\left| \langle 0 |\big(S_{n_-}^{\dagger (Q_{M})} S_{n_+}^{(Q_{M})}\big)(0)\,|0\rangle\right| \equiv R_{c}^{(Q_{M})}  R_{\bar{c}}^{(Q_{M})} \;,
\label{eq:softsubtractionapp}
\end{equation}
where $Q_M$ is the charge of an outgoing meson $M$ with $\bar{u}$ and $q_1$ quarks, and $c (\bar{c})$ denote the collinear (anti-collinear) directions. The matrix element in \eqref{eq:softsubtractionapp} equals unity to all orders in $\alem$ in dimensional regularization, because the soft Wilson lines generate only scaleless integrals. For consistency, we proceed by computing  the soft matrix element with the same dimensional UV and off-shell IR regularization as used to calculate \eqref{eq:pionZfactor}. During the soft decoupling, the soft Wilson lines $S_{n_\pm}$ inherit the off-shellness of their associated hard-collinear field, which changes the Wilson line propagators (see also \cite{Beneke:2020vnb} and Appendix A \cite{Beneke:2019slt}). For an incoming photon with momentum $k$, the propagator of the $S^\dagger_{n_-}$ Wilson line becomes
\begin{equation}
\label{eq:offshellness}
   \frac{1}{n_- k - i 0^+}  \to \frac{1}{n_- k -{\delta}_c - i 0^+} \ ,
\end{equation}
where the quantity $\delta_c\equiv k_{\bar{u}}^2/(n_+k_{\bar{u}}) =  k_{q}^2/(n_+k_{q}) $ is the same\footnote{This follows from the identity $S^{\dagger (d)}_{n_-} S^{(u)}_{n_-} =S^{\dagger (Q_M)}_{n_-}$.} for both the $\bar{u}$ and $q$ quarks of the meson $M$. For the $S_{n_+}$ Wilson line, we exchange $n_- \to n_+$ and $\delta_c \to \delta_{\bar{c}}  $.  
In order to explicitly calculate the matrix element, we expand the soft Wilson lines and obtain the one-loop correction
\begin{align}
 \int &\frac{d^dk}{(2\pi)^d}  \frac{-2i}{k^2+i0}  \frac{Q_M e}{n_- k - \delta_c - i0}  \frac{Q_{M} e}{n_+ k + \delta_{\bar{c}} +i0} \ .
\end{align}
Evaluating the above integral gives
\begin{align}
\langle 0 |(S_{n_-}^{\dagger (Q_{M})} S_{n_+}^{(Q_{M})})(0)\,|0\rangle & =  1-   \frac{\alem}{4\pi} Q_M^2 \left[\frac{2}{\epsilon^2}  + \frac{2}{\epsilon} \ln{\frac{\mu}{-\delta_c}} +  \frac{2}{\epsilon}\ln{\frac{\mu}{\delta_{\bar{c}}}} 
\right. \nonumber  \\
   &{}   \left. +  \ln^2{\frac{\mu}{-\delta_c}}  + \ln^2{\frac{\mu}{\delta_{\bar{c}}}}   +  2\ln{\frac{\mu}{-\delta_c}}  \ln{\frac{\mu}{\delta_{\bar{c}}}} + \frac{\pi^2}{2}  \right]  .
    \label{eq:ssmatapp0}
\end{align}
We assume $\delta_{c,\bar{c}} < 0$ and the $i0$-prescription is implicitly given by replacing $\delta_{c,\bar{c}} \to \delta_{c,\bar{c}}+i0$. Hence, imaginary parts arise when rewriting $\ln(\mu/(\delta_{\bar{c}} + i0)) = \ln(\mu/(-\delta_{\bar{c}} - i0)) - i \pi$:
\begin{align}
    \langle 0 |(S_{n_-}^{\dagger (Q_{M})}& S_{n_+}^{(Q_{M})})(0)\,|0\rangle  =  1-   \frac{\alem}{4\pi} Q_M^2 \left[\frac{2}{\epsilon^2} +  \frac{2}{\epsilon}\ln{\frac{\mu}{-\delta_{c}}}  + \frac{2}{\epsilon} \ln{\frac{\mu}{-\delta_{\bar{c}}}}  -\frac{2i \pi}{\epsilon} 
\right. \nonumber  \\
   &{}   \left. + \ln^2{\frac{\mu}{-\delta_{c}}}   +  \ln^2{\frac{\mu}{-\delta_{\bar{c}}}}   +  2  \ln{\frac{\mu}{-\delta_c}} \ln{\frac{\mu}{-\delta_{\bar{c}}}}  - \frac{\pi^2}{2} 
  - 2i\pi \left( \ln{\frac{\mu}{-\delta_c}}+ \ln{\frac{\mu}{-\delta_{\bar{c}}}} \right) \right]  .
    \label{eq:ssmatapp}
\end{align}
We note that the subtraction factors $R_{c}$ and $R_{\bar{c}}$ are defined using the absolute value of the soft matrix elements exactly to avoid imaginary parts, arising from soft rescattering phases, in the collinear sector. 

The matrix element in \eqref{eq:ssmatapp} can now be separated into $R_c$ and $R_{\bar{c}}$. For the divergent parts, we define this split such that $R_c$ only depends on $\delta_c$ and equivalently such that $R_{\bar{c}}$ only depends on $\delta_{\bar{c}}$. The finite terms are split such that $R_{\bar{c}}$ can be obtained from $R_c$ by switching $n_-\leftrightarrow n_+$. However, as the finite terms depend both on $\delta_c$ and $\delta_{\bar{c}}$, this leaves an ambiguity which we resolve by explicitly defining
\begin{align}
R_{{c}}^{(Q_{M})} & =  1 -  \frac{\alem}{4\pi} Q_{M}^{\,2} \left[\frac{1}{\epsilon^2} + \frac{2}{\epsilon}\ln{\frac{\mu}{-\delta_{ c}}}  + \ln^2{\frac{\mu}{-\delta_{c}}}   +  \ln{\frac{\mu}{-\delta_c}}  \ln{\frac{\mu}{-\delta_{\bar{c}}}}  - \frac{\pi^2}{4} \right] , \\
R_{\bar{c}}^{(Q_{M})} & = 1 -  \frac{\alem}{4\pi} Q_{M}^{\,2} \left[\frac{1}{\epsilon^2} + \frac{2}{\epsilon}\ln{\frac{\mu}{-\delta_{\bar c}}} +  \ln^2{\frac{\mu}{-\delta_{\bar{c}}}}    +   \ln{\frac{\mu}{-\delta_c}}  \ln{\frac{\mu}{-\delta_{\bar{c}}}} - \frac{\pi^2}{4} \right]  .
\end{align}


\section{Details on Mellin integrals near the endpoints}

In Section~\ref{subsec:pionEP}, we found that the linear endpoint behaviour of the LCDA $\Phi_M(u;\mu)$ for small $u$ is modified by a non-integer power of logarithms $\ln u$, more precisely $\Phi_M(u;\mu) \sim u(-\ln u)^p$, with \mbox{$p = \alem Q_{q_1} Q_M/(\alpha_s C_F + \alem Q_{q_1} (Q_{q_2}-Q_M))$}. We mentioned this explicitly in the main text in QCD$\times$QED with renormalization-scale independent gauge couplings. In this appendix, we prove our statement that this result also holds in the general case of running coupling constants, provided that $\alpha_s(\mu)$ and $\alem(\mu)$ are evaluated at the scale $\mu$ in the exponent $p = p(\mu)$. Moreover, we add details on the analytic structure and how to perform the inverse Mellin transform, starting again with the QCD-only case in Appendix~\ref{app:InverseMellinQCD}. 

In Section~\ref{subsec:pionEP}, we argued that the small-$u$ behaviour of $\Phi_M(u;\mu)$ is fully determined by the asymptotic evolution kernel in the soft limit, and that the collinear part of the kernel contributes at most subleading logarithms $\ln u$. If $\Phi_M(u;\mu) \sim u^b$ with $b = 1$ the convolution with the soft asymptotic kernel becomes UV divergent. We emphasize again, that in the expansion by regions the power-counting of the integrand determines the small-$u$ behaviour, independent of the (non-)convergence of the integrals. Nevertheless, the actual computation of the integrals requires introducing an additional regulator. The dependence on the regulator cancels only after including the collinear region in the evolution kernel. In this appendix, to keep the mathematical expressions as simple as possible, but without loss of generality, we choose a simple power-like initial condition with an explicit upper cut-off $\Lambda \sim \mathcal{O}(1)$, 
$\hat{\Phi}_M(u;\mu_0) =\theta(\Lambda-u) u^b$, for which we can calculate the endpoint behaviour explicitly. This choice of cut-off has the advantage that the Mellin transform
\begin{align} 
\label{eq:softinicond}
\tilde{\hat{\Phi}}_M(\eta;\mu_0) = \frac{\Lambda^{b-\eta}}{b-\eta}
\end{align}
yields only a single simple pole at $\eta = b$ in the complex $\eta$-plane. The Mellin integral converges for ${\rm Re}(\eta)<b$, so that $c<b$ must be chosen for the inverse.

\subsection{Inverse Mellin transform in QCD}
\label{app:InverseMellinQCD}

In QCD-only, the solution to the soft RGE in Mellin space is now given by \eqref{eq:mellinsolutionQCD} together with~\eqref{eq:softinicond}: 
\begin{align}
 \phi_M(u;\mu) = \Lambda^{b_\mu} \,e^{(2\gamma_E-3/2)a} \int_{c-i\infty}^{c+i\infty} \frac{d\eta}{2\pi i} \left( \frac{u}{\Lambda} \right)^\eta \frac{\Gamma(1-\eta) \Gamma(1+\eta+a) }{\Gamma(1+\eta)\Gamma(1-\eta-a)} \, \frac{1}{b_\mu-\eta} \,.
\end{align}
The evolution variable $a$ is defined in \eqref{eq:aQCD}, and $b_\mu \equiv b - a(\mu,\mu_0)$. In the complex $\eta$-plane, the integrand is analytic up to strings of simple poles from the gamma functions in the numerator and one pole from the Mellin transformed initial condition. The two strings of poles are located at $\eta=-1-a-n$ and $\eta=1+n$, where $n = 0,1,2,\dots $ is a non-negative integer, and the single pole from the initial condition is located at $\eta=b_\mu$. In the limit $u\to 0$ we can safely assume $u<\Lambda$. The factor $(u/\Lambda)^\eta$ then exponentially suppresses the integrand for $\rm{Re}(\eta)\to \infty$, such that we can close the integration contour in the right half-plane. The asymptotic behaviour of the gamma functions for large real arguments requires us to choose $c$ in the interval $-1-a < c < \min(1,b_\mu)$. This implies the constraints $b>-1$ and $a>-2$, which excludes an overlap between the two strings of poles.\footnote{Evolution over a very large scale interval would lead to 
$a<-2$. In this case, the evolution can be done in smaller 
serial intervals, or through a properly defined analytic continuation. This is never relevant to physical processes.}
After deformation, the new contour encircles the poles at $\eta=b_\mu,1+n$ in the mathematically negative direction, and the contour integral is given by the sum of all residues at these poles. This in particular implies that the left-most pole satisfying ${\rm Re}(\eta)>c$  gives the dominant contribution in the limit $u \to 0$:
\begin{align}
   \phi_M(u;\mu) &= e^{(2\gamma_E-3/2)a} \Lambda^{b_\mu} \sum_{\eta = b_\mu,1+n} (-1)\cdot{\rm Res}\bigg[ \left( \frac{u}{\Lambda} \right)^\eta \frac{\Gamma(1-\eta) \Gamma(1+\eta+a)}{\Gamma(1+\eta)\Gamma(1-\eta-a)} \, \frac{1}{b_\mu-\eta} \bigg]\nn \\&=  e^{(2\gamma_E-3/2)a} \left(  \frac{\Gamma(1+b)\Gamma(1-b_\mu)}{\Gamma(1-b)\Gamma(1+b_\mu)} \,\,u^{b_\mu} +  \frac{\Lambda^{b_\mu-1} }{b_\mu-1}\frac{\Gamma(2+a)}{\Gamma(-a)} u + \mathcal{O}(u^2) \right) .
\end{align}
The first term is independent of the regulator $\Lambda$. This is expected as for $b_\mu<1$ the collinear region is power-suppressed, and the endpoint behaviour at the scale $\mu$ is fully determined by the endpoint-behaviour of the initial condition at the scale $\mu_0$. The regulator dependence in the linear term, however, is cancelled only after including the collinear region in the evolution. We note that for $b_\mu \to 1$ the sum of both terms is finite. 

Hence, depending on whether $b_\mu$ is smaller or larger than one, the endpoint behaviour will be either linear or $ u^{b_\mu}$. Since we have $a<0$ in QCD, we can conclude that the RG solution always flows towards a linear endpoint behaviour. The same analysis holds for the endpoint $u=1$, showing consistency with the asymptotic form of $\phi_M(u;\mu \to \infty) \to 6 u\bar{u}$.
 
\subsection{Inverse Mellin transform in QCD$\times$QED} 
\label{subsec:contourint}
We now turn to the general case of QCD$\times$QED and show how additional non-integer powers of logarithms $\ln u$ arise from the contour integral. As the evolution above and below the critical scale $\mu_c$ defined in \eqref{eq:mucrit} yields slightly different results, we analyze them separately. We start with the more relevant case $\mu < \mu_c$, which implies that the evolution variable $a(\mu,\mu_0)<0$ is negative for all scales with $\mu_0 < \mu < \mu_c$. In QCD$\times$QED, the analysis of the soft region is more involved due to the additional exponential of the integral over harmonic number functions in \eqref{eq:mellinsolution}. With the initial condition~\eqref{eq:softinicond}, \eqref{eq:mellinsolution} reads 
\begin{align}\label{eq:QEDmellinapp}
\tilde{\hat{\Phi}}_M(\eta;\mu) &= \exp\left[2 \gamma_E a -3\check{a}/2\right]\; \frac{\Gamma(1-\eta) \Gamma(1+\eta+a)}{\Gamma(1+\eta)\Gamma(1-\eta-a)} \, \frac{\Lambda^{b_\mu-\eta}}{b_\mu-\eta} \nonumber \\
&\times \exp\left\{ -\int_{\mu_0}^\mu \frac{d\mu'}{\mu'} \frac{\alpha_{\rm em}(\mu') Q_{q_1}Q_M}{\pi} \left( H_{\eta + a(\mu,\mu')} + H_{-\eta-a(\mu,\mu')} \right) \right\} \,,
\end{align}
where $b_\mu$ is defined in QCD$\times$QED by~\eqref{eq:adef}. The analytic structure of this exponential is non-trivial, since the integration variable $\mu'$ appears in the harmonic number function argument. The harmonic number functions have simple poles located at the negative integers, and consequently the integral has branch cuts on the real $\eta$-axis, as discussed below. As in QCD-only, the parameter $c$ for the inverse transformation must be chosen in interval $-1-a< c< \min(1,b_\mu)$, separating the cuts into left (due to $H_{\eta+a(\mu,\mu')}$) and right contributions (due to $H_{-\eta-a(\mu,\mu')}$). Asymptotically, the harmonic numbers behave like $\ln \eta$ for large real arguments, such that for $u\to 0$, we can still deform the contour to enclose the discontinuities on the right-hand side. 
Again, the left-most pole or cut for ${\rm Re}(\eta) > c$ then determines the endpoint behaviour. For $b_\mu<1$, we just pick up the residue at $\eta=b_\mu$ as in QCD-only and the resulting endpoint behaviour is $u^{b_\mu}$. Once reaching $b_\mu=1$, we need to analyze the solution~\eqref{eq:QEDmellinapp} for the initial condition $b=1$ as it is a stable point of the evolution. 
At $\eta=1$, the exponential factor in the second line of \eqref{eq:QEDmellinapp} has a branch-point due to the left-most pole of the harmonic number function, $H_{-\eta-a(\mu,\mu')} = -1/(1-\eta-a(\mu,\mu'))~ +$ regular terms and poles/cuts further to the right. The relevant non-analytic term can be extracted from 
\begin{eqnarray}
\label{eq:lowestbranchcut}
    -\int_{\mu_0}^\mu \frac{d\mu'}{\mu'} \, \frac{\alem(\mu') Q_{q_1}Q_M}{\pi}  H_{-\eta-a(\mu,\mu')} &=& \int_{\mu_0}^\mu \frac{d\mu'}{\mu'} \, \frac{\alem(\mu') Q_{q_1}Q_M}{\pi} \frac{1}{1-\eta-a(\mu,\mu')}\nn \\ 
&&\,\hspace*{-4cm}+ \text{ regular terms and poles/cuts further to the right} \; .
\end{eqnarray}
In the integration domain in $\mu'$, the function $a(\mu,\mu')$ takes values in the interval $[a,0]$, with $a = a(\mu,\mu_0) <0$. For given $\mu$, the integral in~\eqref{eq:lowestbranchcut} thus has a branch-cut of finite length, starting at $\eta = 1$ and extending to $\eta=1-a$. We can conclude in general that the harmonic number function leads to cuts on the intervals $[1+n,1-a+n]$ with $n=0,1,2,\dots$ a non-negative integer. 
To determine the contribution from the left-most branch-cut, we recall that the branch-point at $\eta = 1$ requires $\mu' \to \mu$, such that we can expand
\begin{align} \label{eq:aexp}
    a(\mu,\mu') = -\frac{\alpha_s(\mu) C_F + \alem(\mu) Q_{q_1} (Q_{q_2} - Q_M)}{\pi} \, \ln \frac{\mu}{\mu'} + \ldots 
\end{align}
in the denominator of the integrand in~\eqref{eq:lowestbranchcut}, and replace $\alem(\mu') \to \alem(\mu)$ in the numerator. The integral in $\mu'$ then essentially reduces to the scale-independent case already discussed in the main text: 
\begin{eqnarray} \label{eq:harmonicExp}
   -\int_{\mu_0}^\mu \frac{d\mu'}{\mu'} \, \frac{\alem(\mu') Q_{q_1}Q_M}{\pi}  H_{-\eta-a(\mu,\mu')}  &=& p(\mu) \ln \frac{1}{1-\eta} \nn \\
&& \hspace*{-4.5cm} 
+ \, \text{regular terms and poles/cuts further to the right},
\end{eqnarray}
but with $\mu$-dependent $p(\mu) = \alem(\mu) Q_{q_1}Q_M /( \alpha_s(\mu) C_F + \alem(\mu) Q_{q_1}(Q_{q_2} - Q_M))$. The function $p(\mu)>0$ is always positive for evolution below $\mu_c$; at $\mu = \mu_c$, 
 it is singular and flips its sign. The result~\eqref{eq:harmonicExp} for the exponential needs to be combined with the pole from the Laurent expansion of $\Gamma(1-\eta)$ in~\eqref{eq:QEDmellinapp}. The regular and other singular terms are irrelevant for the asymptotic small-$u$ behaviour, and can thus be absorbed, along with other irrelevant factors, into a constant $\kappa$. The inverse 
Mellin transformation now takes the form 
\begin{align}
\hat{\Phi}_M(u;\mu) &= \kappa  \int_C \frac{d\eta}{2\pi i} \,u^\eta \left(\frac{1}{1-\eta}\right)^{1+p(\mu)} + \mathcal{O}(u^2)  \, .
\end{align}
where the contour $C$ encloses the cut for ${\rm Re}(\eta)>1$, as discussed above.
According to \cite{Beneke:2016jpx}, the contour $C$ must be carefully chosen. We decompose it into a small circle $C_\varepsilon$ of radius $\varepsilon$ around $\eta=1$, that is parametrized by $\eta=1+\varepsilon e^{i\varphi}$ for $\varphi \in (2\pi,0)$, and the contour $C_{\rm cut}$ consisting of straight lines above and below the real axis extending from $1+\varepsilon$ to $\infty$. The contributions from the cut and the circle are given by
\begin{align}
    \hat{\Phi}^{C_{\rm cut}}_M(u;\mu) &= \kappa \int_{1+\varepsilon}^\infty \frac{d\eta}{2\pi i} \,u^\eta~{\rm disc} \bigg[\left(\frac{1}{1-\eta}\right)^{\!1+p} \bigg] =  \kappa \,\frac{u(-\ln u)^p}{\Gamma(1+p)\Gamma(-p)}\Gamma(-p,-\varepsilon \ln u) \; , \label{eq:phicut}\\
    \hat{\Phi}^{C_\varepsilon}_M(u;\mu) &= \kappa\, \frac{ u (-\ln u)^p}{\Gamma(1+p)\Gamma(-p)} \sum_{n=0}^\infty \frac{(-1)^n (-\varepsilon \ln u)^{n-p}}{n!(n-p)} \; ,\label{eq:phieps}
\end{align}
respectively, where the discontinuity is defined as ${\rm disc}(f(\eta)) = f(\eta+i0)-f(\eta-i0)$ and $\Gamma(-p,-\varepsilon\ln u)$ is the incomplete gamma function
\begin{align}
    \Gamma(-p,-\varepsilon \ln u) = \Gamma(-p) - \sum_{n=0}^\infty \frac{(-1)^n (-\varepsilon \ln u)^{n-p}}{n!(n-p)} \; .
\end{align}
For $p(\mu)>0$, the first terms $n<p$ of the sum are divergent as $\varepsilon \to 0$, but they cancel exactly after adding \eqref{eq:phicut} and \eqref{eq:phieps}. In total, we obtain
\begin{align} \label{eq:qedlnp}
\hat{\Phi}_M(u;\mu) &=   \hat{\Phi}^{C_{\rm cut}}_M(u;\mu)+\hat{\Phi}^{C_\varepsilon}_M(u;\mu) + \mathcal{O}(u^2)=\frac{\kappa}{\Gamma(1+p(\mu))} \,u(-\ln u)^{p(\mu)} + \mathcal{O}(u^2) \, ,
\end{align}
which is also dictated by Watson's lemma. This confirms our statement in~\eqref{eq:lnp} for scale-dependent gauge couplings.

The discussion changes when evolution above $\mu_c$ is considered. For $\mu>\mu_0>\mu_c$ the evolution variable $a(\mu,\mu_0)$ becomes positive and hence the cut from the harmonic number function starts at $\eta=1-a$. The relevant contribution in \eqref{eq:QEDmellinapp} is then either given by the point $\eta=b_\mu$ or $\eta=1-a$. In principle, the latter also gives rise to $\ln u$-modified terms from the harmonic numbers. However, this contribution will never be physically relevant since, independent of the value of $b$, evolution will 
immediately generate the endpoint behaviour $u^{1-a(\mu_0+d\mu,\mu_0)}$ with $1-a(\mu_0+d\mu,\mu_0)<1$. This justifies to assume $b<1$ from the beginning, and consequently $b_\mu<1-a<1$, such that the dominant contribution from the inverse transformation of~\eqref{eq:QEDmellinapp} is always given by the residue at $\eta = b_\mu$. We conclude that, above the critical scale, the power-like behaviour is always proportional to $u^{b_\mu}$ in QED. Hence, at some large but finite $\mu$, the evolution will run to solutions with endpoint behaviour $u^{-1}$ 
($b_\mu=-1$), for which the RGE is no longer defined.


\bibliographystyle{JHEP} 
\bibliography{refs.bib}
\end{document}